\documentclass{aastex}
\usepackage{spr-astr-addons}
\usepackage{url}
\usepackage{graphicx}
\usepackage{epsfig}
\usepackage{epstopdf}
\usepackage{latexsym}
\usepackage{graphics}
\usepackage{color} 

\RequirePackage{color}

\begin{document}

\title{Nuclear structure and weak  rates
of heavy waiting point nuclei under $rp$-process conditions}

\shorttitle{Nuclear structure and weak  rates of heavy waiting point
nuclei under $rp$-process conditions} \shortauthors{Nabi and
B\"{o}y\"{u}kata}

\author{Jameel-Un Nabi\altaffilmark{1}} \and \author{Mahmut B\"{o}y\"{u}kata\altaffilmark{2}}
\altaffiltext{1}{Faculty of Engineering Sciences,\\GIK Institute of
Engineering Sciences and Technology, Topi 23640, Khyber Pakhtunkhwa,
Pakistan.}
\altaffiltext{2}{Physics Department, Science and Arts Faculty, K\i r\i kkale University, 71450, K\i r\i
kkale, Turkey}
\altaffiltext{1}{jameel@giki.edu.pk}

\begin{abstract}
The structure and the weak interaction mediated rates of the heavy
waiting point (WP) nuclei $^{80}$Zr, $^{84}$Mo, $^{88}$Ru, $^{92}$Pd
and $^{96}$Cd along  $N = Z$ line were studied within the
interacting boson model-1 (\mbox{IBM-1}) and the proton-neutron
quasi-particle random phase approximation (\mbox{pn-QRPA}). The
energy levels of the $N$ = $Z$ WP nuclei were calculated by fitting
the essential parameters of \mbox{IBM-1} Hamiltonian and their
geometric shapes were predicted by plotting potential energy
surfaces (PESs). Half-lives, continuum electron capture rates,
positron decay rates, electron capture cross sections of WP nuclei,
energy rates of $\beta$-delayed protons and their emission
probabilities were later calculated using the \mbox{pn-QRPA}. The
calculated Gamow-Teller strength distributions were compared with
previous calculation. We present positron decay $\&$ continuum
electron capture rates on these WP nuclei under $rp$-process
conditions using the same model. For the $rp$-process conditions,
the calculated total weak rates are twice the Skyrme HF+BCS+QRPA
rates for $^{80}$Zr. For remaining nuclei the two calculations
compare well. The electron capture rates are significant and compete
well with the corresponding positron decay rates under $rp$-process
conditions. The finding of the present study supports that electron
capture rates form an integral part of the weak rates under
$rp$-process conditions and has an important role for the nuclear
model calculations.
\end{abstract}



\section{\label{sec:level1} Introduction}

Production of heavy nuclei via a series of rapid proton capture
reactions is normally called as the $rp$-process. The process halts
when a radioactive element is produced and then proceeds through a
$\beta$-decay (with a low probability of double proton capture).
Later, the reaction flow has to wait for a comparatively slower
$\beta$-decay and this kind of nucleus is named as a waiting point
(WP). Typical time scale of the $rp$-process and the half-lives of
the WP nuclei is roughly $\sim$ 100 $s$. Thus the time scale of the
nucleosynthesis process and isotopic abundances can be determined.
The weak-interaction mediated stellar rates of the neutron-deficient
heavy nuclei play a significant role to understand the $rp$-process.
As argued in Ref. \citep{Sch98}, for a successful $rp$-process
nucleosynthesis calculation, the nuclear masses and  deformations
are the important parameters for a reliable calculation of the
stellar electron capture and positron decay rates of the WP nuclei
along a given reaction path since the time structure and the
abundance patterns can be described by them.

For the nuclear structure studies, one of the useful models is the
interacting boson model-1 ~\mbox{(IBM-1)}~\citep{Iachello87} which
has connections with the shell model~\mbox{(SM)}~\citep{Talmi93} and
the geometric collective model~\mbox{(GCM)}~\citep{Bohr98} as seen
in Refs.~\citep{Arima80, Iachello87i, Dieperink80a, Dieperink80b,
Ginocchio80a, Ginocchio80b, Isacker81}. The study of the structural
properties of the nuclei within the IBM model is still an active
research area performed in recent years~\citep{Boyukata10,
Nomura11a, Bentley11, Sharrad12, Gupta13, Ramos14a, Sharrad15,
Khalaf15, Sirag15, Nab16}.

The aim of the this work is to investigate some nuclear structural
properties of the WP nuclei within the ~\mbox{IBM-1} and the
proton--neutron quasiparticle random phase approximation~\mbox{(pn-QRPA)} models and to complete investigation

of the WP nuclei along the $N$ = $Z$ line which we previously
started \citep{Nab16}. As an added feature the calculation of the
electron capture cross sections of these WP nuclei was performed for
the first time in this paper. First, the energy levels and the
potential energy surfaces (\mbox{PESs}) of each nucleus along
$N$~=~$Z$ line were studied within the ~\mbox{IBM-1}. For the
calculation of the energy levels, essential Hamiltonian parameters
were fitted from the experimental data \citep{NNDC} for known
$^{80}$Zr, $^{84}$Mo, $^{88}$Ru, $^{92}$Pd nuclei and from the shell
model calculation for unknown $^{96}$Cd nucleus \citep{Coraggio12}.
To define the geometric shapes of WP nuclei, the common Hamiltonian
parameters were used in the \mbox{PESs} formula; $V(\beta,\gamma)$.
Later we used the \mbox{pn-QRPA} model to calculate Gamow-Teller
(GT) strength distributions, terrestrial half-lives, electron
capture cross sections in stellar environment and stellar weak rates
of the WP nuclei. The later includes positron decay rates, continuum
electron capture rates, energy rates of $\beta$-delayed protons and
their emission probabilities.

For the description of decay of unstable nuclei in the direction of
line of stability, the energy rates of $\beta$-delayed protons and
probabilities for $\beta$-delayed proton emission rates are
important prerequisites. These rates are required for a better
understanding of the $rp$-process in Type I X-ray bursts
\citep{Sch06}. The burst ashes constitute a key input parameter for
neutron star crust models \citep{Gup07}. The lifetimes of WP nuclei
determine the composition of these burst ashes. $\beta$-delayed
proton emission rates of nuclei along the $rp$-process path  and the
associated probabilities may alter the composition of the burst
ashes. In literature the data on probabilities of $\beta$-delayed
proton emission is rather scarce. These probabilities influence the
final abundance of heavy elements. In past the $\beta$-delayed
proton emission has been treated in crude approximations. For a
reliable prediction of these weak rates an accurate description of
the $\beta$ strength function is a prerequisite and necessitates a
microscopic calculation of the same (see also \citep{Hir92}).

The outline of the paper as follows. The summary of main theme is
presented in this section. In Section~2, the \mbox{IBM-1} model, the
dynamical symmetries  and the potential energy surface are briefly
described along with the model Hamiltonian and the formalism of
PESs. In Section~3, the \mbox{pn-QRPA} model formalism for the
calculation of the GT strength distributions, the stellar weak rates
and the electron capture cross section are discussed. Section~4
includes the results of all calculations and the comparisons of the
results with the measured data and also with the previous
calculations. Finally, the conclusions are reported in Section~5.

\section{\label{sec:level2} The interacting boson model-1 (IBM-1)}

The \mbox{IBM-1} model is a group theoretical approach firstly
introduced by Fescbach and Iachello~\citep{Fescbach73, Fescbach74}
and later developed by Arima and Iachello~\citep{Iachello74,
Arima75i, Arima75ii, Arima76, Arima78i, Arima78ii, Arima78iii,
Arima79}. The group structure of this model is discussed within the
$U(6)$ group which arises from the s-bosons (with $L=0$, $\mu=0$)
and d-bosons (with $L=2$, $-2\leq$ $\mu$ $\leq2$). The $U(6)$ group
is characterized by the boson numbers $N$.

\subsection{\label{sec:level2i} The dynamical symmetries}

The unitary group $U(6)$ has three possible subgroups; $U(5)$,
$SU(3)$ and $O(6)$. These subgroups are generally termed as
$dynamical$ $symmetries$ and are obtained by constructing chains of
each subgroup of the $U(6)$ group ~\citep{Iachello87}:

    $U(6)\supset$  $U(5)\supset$ $O(5)\supset$ $O(3)$

    $U(6)\supset$  $SU(3)\supset$ $O(3)$

    $U(6)\supset$  $O(6)\supset$ $O(5)\supset$ $O(3)$. \\
Each  dynamical symmetry has a corresponding geometrical
relationship with nuclei. The $U(5)$ symmetry corresponds to
spherical nuclei (spherical harmonic vibrator)
~\citep{Iachello74,Arima76}, the SU(3) symmetry is related to well
deformed nuclei (axially symmetric shapes~\citep{Arima78ii}) and
$O(6)$ to $\gamma$-unstable nuclei (axially asymmetric
shapes)~\citep{Arima78i,Arima79}. Moreover, other nuclei located in
transitional regions (in between these three shapes) can also be
investigated as discussed in Ref.~\citep{Arima78iii}. To get an idea
of the geometric shape of a given nucleus, one can look at its
energy ratio $R_{4/2}$. 
For a given dynamical symmetry, the energy ratio $2.00$, $3.33$ and
$2.50$ corresponds to spherical, axially deformed (prolate or
oblate) and $\gamma$-unstable nuclei, respectively.

\subsection{\label{sec:level2ii} The IBM-1 Hamiltonian}

The simplest form of this model is abbreviated as \mbox{IBM-1} and
in this model protons and neutrons are taken as nucleons.
The \mbox{IBM-1} model is quite useful to describe the
nuclear structural properties of even-even nuclei by
interacting $s$- and $d$- bosons and these interactions
can be used as free parameters in the \mbox{IBM-1} Hamiltonian.
The general Hamiltonian, with six interactions in total,
includes combinations of the bosonic creation and annihilation
operators $s$, $s^{\dag}$, $d$, $d^{\dag}$ and also can be
constructed as the sum of the Casimir operators of
$U(5), SU(3), O(6)$~\citep{Iachello87}.

For the present work, the following \mbox{IBM-1}
Hamiltonian in the multipole form including spherical
and deformed part were used:

\begin{equation}
\hat H= \epsilon\,\hat n_d + a_2 \,\hat Q\cdot\hat Q .
\label{e_s.ham}
\end{equation}
Here, $\hat n_d$ and $\hat Q$ are the boson-number and
the quadrupole operators, respectively, defined as a function of boson operators
\begin{eqnarray}
\hat n_d&=&\sqrt{5}[d^\dag\times\tilde d]^{(0)}_0,
\nonumber\\
\hat Q&=&[d^\dag\times\tilde s+s^\dag\times\tilde d]^{(2)}+
\overline{\chi}[d^\dag\times\tilde d]^{(2)},
\label{e_term}
\end{eqnarray}
where $\overline{\chi}$ is the third parameter of Hamiltonian
(\ref{e_s.ham}). The $\epsilon$, ${a_2}$ and $\overline{\chi}$
parameters can be obtained by fitting to experimental data
~\citep{NNDC} to describe the collective properties of the given
nucleus. For a general Hamiltonian with more parameters see
Ref.~\citep{Iachello87}. To gain more insight into the model
parameters see Ref.~\citep{Casten88}.

\subsection{\label{sec:level2iii} The geometry of nuclei: Potential energy surface (PES)}

The geometric shapes of the nuclei can be predicted within the IBM
by plotting the potential energy surface (PES) as function of
deformation parameters ($\beta$, $\gamma$). This potential energy
surface $V(\beta,\gamma)$ can be obtained from the \mbox{IBM-1}
Hamiltonian in the classical limit~\citep{Dieperink80a,
Dieperink80b, Ginocchio80a, Ginocchio80b}. The PES in the
\mbox{IBM-1} has a common role in the geometric collective model
(GCM) of Bohr and Mottelson~\citep{Bohr98} as following; both
deformation parameters $\beta$ and $\gamma$ are \emph{zero} for the
spherical nuclei, while for the non-spherical nuclei; $\beta\neq0$,
$\gamma=0^\circ$, $30^\circ$, $60^\circ$ for prolate, triaxial, and
oblate shapes, respectively. Moreover, the classical limit for the
Hamiltonian of each $U(5)$, $O(6)$, $SU(3)$ symmetries can also be
derived and their PES can be plotted ~\citep{Dieperink80a,
Isacker81}. However, for present application $V(\beta,\gamma)$,
including spherical and deformed terms, is reduced for the
Hamiltonian (\ref{e_s.ham}) as follows

\tiny{
\begin{equation}
\begin{split}
V(\beta,\gamma)=\epsilon \frac{N\beta^2}{1+\beta^2} + a_2 N\\
\left[\frac{5+(1+\overline{\chi}^2)\beta^2}{1+\beta^2} +(N-1)
\frac{\left(\frac{2\overline{\chi}^2\beta^4}{7}-4\sqrt{\frac{2}{7}}\overline{\chi}\beta^3\cos(3\gamma)+4\beta^2\right)}{(1+\beta^2)^2}\right],
\label{e_s.pes}
\end{split}
\end{equation}
}\normalsize{where $N$ is the boson numbers, $\overline{\chi}$ is
same in the quadrupole operator given by Eq. (\ref{e_term}) and
other constants are common parameters given in the
Hamiltonian~(\ref{e_s.ham}), in which the parameters are adjusted
from experimental data ~\citep{NNDC}. Eq. (\ref{e_s.pes}) includes
the deformation parameters, $\beta$ and $\gamma$, playing the same
role as in the GCM model as discussed above.

\section{\label{sec:level3} The \mbox{pn-QRPA} model formalism}

The proton-neutron residual interactions are taken as particle-hole
and particle-particle interactions in the \mbox{pn-QRPA} formalism
\citep{Mut92}. Here, the particle-hole and particle-particle
interactions are characterized by interaction strength constants
$\chi$ (differentiated from $\overline{\chi}$ introduced above) and
$\kappa$, respectively. The constant $\kappa$ was generally
disregarded in $\beta$-decay calculations \citep{Sta90}, but later
it was understood to be equally important, in particular for the
$\beta^{+}$-decay calculation~\citep{Hir93}. The combination of the
particle-particle force causes to redistribute the calculated
$\beta$ strength shifting commonly toward the lower excitation
energies \citep{Hir93}.

In the present formalism, by using a schematic separable GT residual
forces, the QRPA matrix equation was reduced to an algebraic
equation of fourth order. Solution of this equation was much easier
than the full diagonalization of the non-Hermitian matrix of large
dimensionality \citep{Mut92,Hom96}. Initially a quasiparticle basis,
defined by a Bogoliubov transformation with a pairing interaction
was constructed. Later we solved the RPA equation with a schematic
separable GT residual interaction. At the beginning, single-particle
energies and wave functions were calculated in the Nilsson model
taking into account nuclear deformation. The transformation from the
spherical to the axial-symmetric deformed basis was done using

\begin{equation}\label{sd}
d^{\dagger}_{m\alpha}=\Sigma_{j}D^{m\alpha}_{j}s^{\dagger}_{jm}.
\end{equation}
Here, the $D^{m\alpha}_{j}$ matrices are defined by diagonalizing
the Nilsson Hamiltonian and both particle creation operators are
denoted by $d^{+}$ and $s^{+}$ in the deformed and spherical basis,
respectively. We applied the BCS calculation in the deformed
Nilsson basis for neutrons and protons separately. We employed a
constant pairing force and introduced a quasiparticle basis via

\begin{equation}\label{qbas}
a^{\dagger}_{m\alpha}=u_{m\alpha}d^{\dagger}_{m\alpha}-v_{m\alpha}d_{\bar{m}\alpha}
\\ \nonumber
a^{\dagger}_{\bar{m}\alpha}=u_{m\alpha}d^{\dagger}_{\bar{m}\alpha}+v_{m\alpha}d_{m\alpha},
\end{equation}
where $a^{\dagger}/a$ are the quasiparticle creation/annihilation
operators entering the RPA equation and $\bar{m}$ is the time
reversed state of $m$. The occupation amplitudes $u$ and $v$,
determined from the BCS equations, satisfy the condition $u^{2} +
v^{2} = 1$.

The calculation of stellar weak interaction rates for 709 nuclei
with mass number $A = 18$ to $100$ was first performed within the
\mbox{pn-QRPA} model by Nabi and Klapdor-Kleingrothaus
\citep{Nab99}. Subsequently, a detailed calculation of stellar weak
interaction rates over a wide range of temperature and density scale
for sd- and fp/fpg-shell nuclei were presented in Refs.
\citep{Nab99a,Nab04, Nab99}. These also included the weak
interaction rates for nuclei with mass number $A = 40$ to $44$ (not
yet calculated by shell model). Since then these calculations were
further refined with use of more efficient algorithms, computing
power, incorporation of latest data from mass compilations and
experimental values, and fine-tuning of model parameters both in the
sd-  \citep{Nab07, Nab08, Nab08a} and fp-shell \citep{Nab16, Nab16a,
Nab16b, Nab15, Nab15a, Nab15b, Nab14, Nab14a, Nab12} region. All
theoretical calculations of stellar weak interactions possess
uncertainties. Ref. \citep{Nab08b} discussed in length the
uncertainties associated with the \mbox{pn-QRPA} model and Ref.
\citep{Nab04} discussed in detail the reliability of the
\mbox{pn-QRPA} calculation. Measured data (half lives and
B(GT$_{\pm}$) strength) of thousands of nuclide were compared with
the \mbox{pn-QRPA} calculation and were found to be in reasonable
agreement \citep{Nab04}. In Ref. \citep{Sta90}, the systematic
\mbox{pn-QRPA} calculation of the terrestrial half-lives of
$\beta^{-}$ decays was performed for about 6000 neutron-rich nuclei
between the beta stability line and the neutron drip line. Similar
calculation of the terrestrial half-lives for $\beta^{+}$ decays and
electron capture (EC) for neutron-deficient nuclei with atomic
numbers $Z = 10 - 108$ up to the proton drip line for more than 2000
nuclei were reported in Ref. \citep{Hir93}. These calculations
showed a very good agreement with the existing experimental data
(within a factor of two for more than 90$\%$ (73$\%$) of nuclei with
experimental half-lives shorter than 1 s for $\beta^{-}$
($\beta^{+}$/EC) decays). Many exotic nuclei far from stability have
astrophysical importance and one has to rely on the estimation of
their beta decay properties calculated within theoretical models.
The precision of the \mbox{pn-QRPA} model increases with increasing
distance from the $\beta$-stability line with shorter half-lives
\citep{Sta90, Hir93}. This feature further encourages the use of
pn-QRPA model for the prediction of experimentally unknown
half-lives (specially those present in the stellar interior).

\subsection{\label{sec:level4i} Construction of excited states and nuclear matrix elements}

Once we make transition from terrestrial to stellar conditions, we
need to construct parent and daughter excited states and calculate
the nuclear matrix elements connecting the initial and final states
via the GT$^{+}$ operator (that converts a proton into a neutron)
within the \mbox{pn-QRPA} model. Excited states in this model are
constructed as phonon-correlated multi-quasi-particle states. The
RPA can be formulated for excitations from the $J^{\pi} = 0^{+}$
ground state of a given even-even nucleus. The excited states of
this nucleus are obtained by exciting one-proton (or one-neutron)
and can be described by adding two-proton (or two-neutron)
quasiparticle (q.p.) state to the ground state \citep{Mut92}. The
excited states of our chosen  $N$ = $Z$ WP nuclei are two-proton and
two-neutron q.p. states. Transitions from these initial states are
possible to final proton-neutron q.p. pair states in the odd-odd
daughter nucleus. The transition amplitudes and their reduction to
correlated ($c$) one-q.p. states are given by}

\footnotesize{
\begin{eqnarray}
<p^{f}n_{c}^f \mid t_{\pm}\sigma_{-\mu} \mid p_{1}^{i}p_{2c}^{i}> \nonumber \\
 = -\delta (p^{f},p_{2}^{i}) <n_{c}^{f} \mid t_{\pm}\sigma_{-\mu} \mid p_{1c}^{i}>
+\delta (p^{f},p_{1}^{i}) <n_{c}^{f} \mid t_{\pm}\sigma_{-\mu} \mid
p_{2c}^{i}>
 \label{first}
\end{eqnarray}
}
\begin{eqnarray}
<p^{f}n_{c}^f \mid t_{\pm}\sigma_{\mu} \mid n_{1}^{i}n_{2c}^{i}> \nonumber \\
 = +\delta (n^{f},n_{2}^{i}) <p_{c}^{f} \mid t_{\pm}\sigma_{\mu} \mid n_{1c}^{i}>
-\delta (n^{f},n_{1}^{i}) <p_{c}^{f} \mid t_{\pm}\sigma_{\mu} \mid
n_{2c}^{i}>,
\end{eqnarray}
\normalsize{where $\vec{\sigma}$ is the spin operator, $\mu$ = -1,
0, 1, are the spherical components of the spin operator and
$t_{\pm}$ stands for the isospin raising (lowering) operator. Other
symbols have their usual meaning.

For the daughter (odd-odd) nucleus the ground state is assumed to be
a proton-neutron q.p. pair state of smallest energy. The low-lying
states in an odd-odd nucleus are expressed in the q.p. picture by
proton-neutron pair states (two-q.p. states) or by states which are
obtained by adding two-proton or two-neutron q.p.'s (four-q.p.
states) \citep{Mut92}. Therefore, the states in daughter nucleus are
expressed in q.p. transformation by two-q.p. states (proton-neutron
pair states) or by four-q.p. states (two-proton or two-neutron q.p.
states). Reduction of two-q.p. states into correlated ($c$) one-q.p.
states is given as}

\footnotesize{
\begin{eqnarray}
<p_{1}^{f}p_{2c}^{f} \mid t_{\pm}\sigma_{\mu} \mid p^{i}n_{c}^{i}> \nonumber\\
= \delta(p_{1}^{f},p^{i}) <p_{2c}^{f} \mid t_{\pm}\sigma_{\mu} \mid
n_{c}^{i}> - \delta(p_{2}^{f},p^{i}) <p_{1c}^{f} \mid
t_{\pm}\sigma_{\mu} \mid n_{c}^{i}>
\end{eqnarray}
\scriptsize{
\begin{eqnarray}
<n_{1}^{f}n_{2c}^{f} \mid t_{\pm}\sigma_{-\mu} \mid p^{i}n_{c}^{i}> \nonumber\\
= \delta(n_{2}^{f},n^{i}) <n_{1c}^{f} \mid t_{\pm}\sigma_{-\mu} \mid
p_{c}^{i}> - \delta(n_{1}^{f},n^{i}) <n_{2c}^{f} \mid
t_{\pm}\sigma_{-\mu} \mid p_{c}^{i}>,
\end{eqnarray}
}
\normalsize{
while the four-q.p. states are simplified as}

\scriptsize{
\small{
\begin{equation}
\begin{split}
<p_{1}^{f}p_{2}^{f}n_{1}^{f}n_{2c}^{f} \mid
t_{\pm}\sigma_{-\mu} \mid p_{1}^{i}p_{2}^{i}p_{3}^{i}n_{1c}^{i}>
\\
=\delta (n_{2}^{f},n_{1}^{i})[ \delta (p_{1}^{f},p_{2}^{i})\delta
(p_{2}^{f},p_{3}^{i})
<n_{1c}^{f} \mid t_{\pm}\sigma_{-\mu} \mid p_{1c}^{i}>\\
-\delta (p_{1}^{f},p_{1}^{i}) \delta (p_{2}^{f},p_{3}^{i})
<n_{1c}^{f} \mid t_{\pm}\sigma_{-\mu} \mid p_{2c}^{i}>\\ +\delta
(p_{1}^{f},p_{1}^{i}) \delta (p_{2}^{f},p_{2}^{i}) <n_{1c}^{f} \mid
t_{\pm}\sigma_{-\mu} \mid p_{3c}^{i}>]\\  -\delta
(n_{1}^{f},n_{1}^{i})[ \delta (p_{1}^{f},p_{2}^{i})\delta
(p_{2}^{f},p_{3}^{i})
<n_{2c}^{f} \mid t_{\pm}\sigma_{-\mu} \mid p_{1c}^{i}>\\
-\delta (p_{1}^{f},p_{1}^{i}) \delta (p_{2}^{f},p_{3}^{i})
<n_{2c}^{f} \mid t_{\pm}\sigma_{-\mu} \mid p_{2c}^{i}> \\+\delta
(p_{1}^{f},p_{1}^{i}) \delta (p_{2}^{f},p_{2}^{i} <n_{2c}^{f}\mid
t_{\pm}\sigma_{-\mu} \mid p_{3c}^{i}>]
\end{split}
\end{equation}
}

\small{
\begin{eqnarray}
<p_{1}^{f}p_{2}^{f}p_{3}^{f}p_{4c}^{f} \mid t_{\pm}\sigma_{\mu} \mid
p_{1}^{i}p_{2}^{i}p_{3}^{i}n_{1c}^{i}>
\nonumber\\
=-\delta (p_{2}^{f},p_{1}^{i}) \delta (p_{3}^{f},p_{2}^{i})\delta
(p_{4}^{f},p_{3}^{i})
<p_{1c}^{f} \mid t_{\pm}\sigma_{\mu} \mid n_{1c}^{i}> \nonumber\\
+\delta (p_{1}^{f},p_{1}^{i}) \delta (p_{3}^{f},p_{2}^{i}) \delta
(p_{4}^{f},p_{3}^{i})
<p_{2c}^{f} \mid t_{\pm}\sigma_{\mu} \mid n_{1c}^{i}> \nonumber\\
-\delta (p_{1}^{f},p_{1}^{i}) \delta (p_{2}^{f},p_{2}^{i}) \delta
(p_{4}^{f},p_{3}^{i})
<p_{3c}^{f} \mid t_{\pm}\sigma_{\mu} \mid n_{1c}^{i}> \nonumber\\
+\delta (p_{1}^{f},p_{1}^{i}) \delta (p_{2}^{f},p_{2}^{i}) \delta
(p_{3}^{f},p_{3}^{i}) <p_{4c}^{f} \mid t_{\pm}\sigma_{\mu} \mid
n_{1c}^{i}>
\end{eqnarray}
}

\small{
\begin{equation}
\begin{split}
<p_{1}^{f}p_{2}^{f}n_{1}^{f}n_{2c}^{f} \mid t_{\pm}\sigma_{\mu} \mid
p_{1}^{i}n_{1}^{i}n_{2}^{i}n_{3c}^{i}>
\\
=\delta (p_{1}^{f},p_{1}^{i})[ \delta (n_{1}^{f},n_{2}^{i})\delta
(n_{2}^{f},n_{3}^{i})
<p_{2c}^{f} \mid t_{\pm}\sigma_{\mu} \mid n_{1c}^{i}>\\
-\delta (n_{1}^{f},n_{1}^{i}) \delta (n_{2}^{f},n_{3}^{i})
<p_{2c}^{f} \mid t_{\pm}\sigma_{\mu} \mid n_{2c}^{i}> \\+\delta
(n_{1}^{f},n_{1}^{i}) \delta (n_{2}^{f},n_{2}^{i}) <p_{2c}^{f} \mid
t_{\pm}\sigma_{\mu} \mid n_{3c}^{i}>] \\-\delta
(p_{2}^{f},p_{1}^{i})[ \delta (n_{1}^{f},n_{2}^{i})\delta
(n_{2}^{f},n_{3}^{i})
<p_{1c}^{f} \mid t_{\pm}\sigma_{\mu} \mid n_{1c}^{i}>\\
-\delta (n_{1}^{f},n_{1}^{i}) \delta (n_{2}^{f},n_{3}^{i})
<p_{1c}^{f} \mid t_{\pm}\sigma_{\mu} \mid n_{2c}^{i}>\\ +\delta
(n_{1}^{f},n_{1}^{i}) \delta (n_{2}^{f},n_{2}^{i}) <p_{1c}^{f} \mid
t_{\pm}\sigma_{\mu} \mid n_{3c}^{i}>]
\end{split}
\end{equation}
}

\small{
\begin{eqnarray}
<n_{1}^{f}n_{2}^{f}n_{3}^{f}n_{4c}^{f} \mid t_{\pm}\sigma_{-\mu}
\mid p_{1}^{i}n_{1}^{i}n_{2}^{i}n_{3c}^{i}>
\nonumber\\
= +\delta (n_{2}^{f},n_{1}^{i}) \delta (n_{3}^{f},n_{2}^{i})\delta
(n_{4}^{f},n_{3}^{i})
<n_{1c}^{f} \mid t_{\pm}\sigma_{-\mu} \mid p_{1c}^{i}> \nonumber\\
-\delta (n_{1}^{f},n_{1}^{i}) \delta (n_{3}^{f},n_{2}^{i}) \delta
(n_{4}^{f},n_{3}^{i})
<n_{2c}^{f} \mid t_{\pm}\sigma_{-\mu} \mid p_{1c}^{i}> \nonumber\\
+\delta (n_{1}^{f},n_{1}^{i}) \delta (n_{2}^{f},n_{2}^{i}) \delta
(n_{4}^{f},n_{3}^{i})
<n_{3c}^{f} \mid t_{\pm}\sigma_{-\mu} \mid p_{1c}^{i}> \nonumber\\
-\delta (n_{1}^{f},n_{1}^{i}) \delta (n_{2}^{f},n_{2}^{i}) \delta
(n_{3}^{f},n_{3}^{i}) <n_{4c}^{f} \mid t_{\pm}\sigma_{-\mu} \mid
p_{1c}^{i}>. \label{last}
\end{eqnarray}
}

\normalsize{
For all the given q.p. transition amplitudes
[Eqs.~(\ref{first})~-~(\ref{last})], the antisymmetrization of the
single-q.p. states was taken into account:\\}

$ p_{1}^{f}<p_{2}^{f}<p_{3}^{f}<p_{4}^{f}$,

$ n_{1}^{f}<n_{2}^{f}<n_{3}^{f}<n_{4}^{f}$,

$ p_{1}^{i}<p_{2}^{i}<p_{3}^{i}<p_{4}^{i}$,

$ n_{1}^{i}<n_{2}^{i}<n_{3}^{i}<n_{4}^{i}$.\\

GT transitions of phonon excitations of each excited state were also
taken into account. We assumed that the quasiparticles in the parent
nucleus remained in the same q.p. orbits. Collective states cannot
be treated in the current \mbox{pn-QRPA} model. To make up for this
deficiency and to further improve the reliability of the calculated
rates, where possible, experimental data were included in the rate
calculation. The calculated excitation energies within the model
were replaced with measured levels when they were within 0.5 MeV of
each other. No theoretical levels were replaced with the
experimental ones beyond the excitation energy for which
experimental compilations without definite spin and/or parity. The
\mbox{pn-QRPA} calculation of the nuclear matrix elements were
described in detail in Ref. \citep{Mut92}.

For the \mbox{pn-QRPA} calculation, one of the most important
parameters is the deformation parameter as argued in
Refs.~\citep{Sta90, Ste04}. Initially, one of the purpose of this
work was to use deformation parameter of the \mbox{IBM-1} model
($\beta_{IBM}$) in the \mbox{pn-QRPA} calculation. However, $\beta$
in IBM model is larger than $\beta$ in geometric  model (GM)
($\beta_{GM}\leq 1.18(2N/A)\beta_{IBM}$ ~\citep{Ginocchio80b}).
According to other work on relation of the $\beta_{IBM}$ model and
the microscopic self-consistent mean-field calculation
~\citep{Nom10}, the $\beta$ parameters in the two models were
different and $\beta_{IBM}$ was larger by a factor $\sim3-5$. As
also discussed in our previous work~\citep{Nab16}, we finally
decided to use the $\beta$ parameter of the relativistic mean-field
(RMF) calculation given in Ref.~\citep{Lal99}. The calculated
$\beta_{IBM}$ and $\beta_{RMF}$ parameters are shown in
Table~\ref{ta1}. The interaction strength parameters $\chi$ and
$\kappa$ were adjusted for the ~\mbox{pn-QRPA} calculation of the GT
strength distributions. Interaction constant $\kappa$ was fixed at
0.1 MeV. On the other hand, the parameter $\chi$ is known to affect
the position of the Gamow-Teller giant resonance and was fitted by
using a $1/A$ dependence \citep{Hir93}. We could conclude a value of
$4.2/A$ as the optimum value of $\chi$ which best reproduced the
experimental half-lives \citep{Aud12,Aud12i} using the deformation
parameter from \citep{Lal99} (see ~\citep{Nab16} for related
discussions). The Q-values were taken from the mass compilation of
Audi et al. \citep{Aud12,Aud12i}. We multiplied results of
\mbox{pn-QRPA} calculated strength by a quenching factor of
$f_{q}^{2}$ = (0.6)$^{2}$ to compare them with the previous
calculations and also with the experimental data, and then to use
them in astrophysical reaction rates.

\subsection{\label{sec:level4ii} Electron capture cross sections}

For low momentum transfer, the nuclear matrix elements of the $\sum
t_{+}\vec{\sigma}$ operator provide the dominant contribution to the
total electron capture cross section \citep{Goo80}. We calculated
these matrix elements using the \mbox{pn-QRPA} model. This
approximation is justifiable at the energy domain prevalent in the
stellar core where energies are low and the energy transfer from the
incident electron to the nucleus is mainly used for nuclear
excitations. The total stellar electron capture cross section on a
target nucleus of mass number $A$ and charge number $Z$, as a
function of incident electron energy $E_{e}$ and stellar temperature
$T$, is given by

\scriptsize{
\begin{eqnarray}
\sigma(E_e,T)=\frac{G_F^{2}cos^2\theta_c}{2\pi}\sum\limits_{i}F(Z,E_e)\frac{(2J_{i}+1)\exp{(-E_i/kT)}}{G(A,Z,T)} \nonumber\\
 \times \sum\limits_{J,f}(E_e-Q+E_i-E_f)^{2}
\frac{|<i|\hat{O}_J|f >|^{2}}{(2J_{i}+1)},
\end{eqnarray}
} \normalsize{where $G_F$ is the Fermi coupling constant, $\theta_c$
is Cabibbo angle, $F(Z,E_e)$ is the well known Fermi function
calculated according to the prescription given in Ref. \citep{Gov71}
and accounts for the distortion of the electron's wave function due
to the Coulomb field of the nucleus. The $Q$ value of the reaction
is the difference in masses of the parent and daughter nuclei. The
$G(A,Z,T)$ are the nuclear partition functions and were calculated
using the recent recipe provided in Ref. \citep{Nab16a, Nab16b}. The
$J_{i}$'s are the angular momenta of parent states. The $\hat{O}_J$
operator, appearing in the nuclear matrix elements of Eq.~(14),
reduces to the GT$^{+}$ operator for low three-momentum transfer. We
further discuss this matrix element below.

\subsection{\label{sec:level4iii} Stellar weak rates}

The positron decay (pd) and continuum electron capture (cec) rates
from the $\mathit{i}$th state of the parent to the $\mathit{j}$th
state of the daughter nucleus is given by

\begin{equation}
\lambda_{ij}^{pd(cec)} =ln2
\frac{f_{ij}^{pd(cec)}(T,\rho,E_{f})}{(ft)_{ij}^{pd(cec)}},
\end{equation}
where $(ft)_{ij}^{pd(cec)}$ is related to the reduced transition
probability $B_{ij}$ of the nuclear transition by

\begin{equation}\label{ft}
(ft)_{ij}^{pd(cec)}=D/B_{ij}.
\end{equation}
The D appearing in Eq.~\ref{ft} is a compound expression of physical
constants,
\begin{equation}
D=\frac{2ln2\hbar^{7}\pi^{3}}{g_{V}^{2}m_{e}^{5}c^{4}},
\end{equation}

\begin{equation}
B_{ij}=B(F)_{ij}+(g_{A}/g_{V})^2 B(GT)_{ij},
\end{equation}
where B(F) and B(GT) are reduced transition probabilities of the
Fermi and ~GT transitions, respectively

\begin{equation}
B(F)_{ij} = \frac{1}{2J_{i}+1} \mid<j \parallel \sum_{k}t_{+}^{k}
\parallel i> \mid ^{2},
\end{equation}

\begin{equation}\label{bgt}
B(GT)_{ij} = \frac{1}{2J_{i}+1} \mid <j
\parallel \sum_{k}t_{+}^{k}\vec{\sigma}^{k} \parallel i> \mid ^{2}.
\end{equation}
For details of calculation of the nuclear matrix elements within the
\mbox{pn-QRPA} formalism we refer to \citep{Mut92}.

The phase space integral $f_{ij}^{pd(cec)}$ is an integral over
total energy,

\begin{equation}\label{ps}
f_{ij}^{pd} = \int_{1}^{w_{m}} w \sqrt{w^{2}-1} (w_{m}-w)^{2} F(-
Z,w) (1-G_+) dw,
\end{equation}
for positron emission, or

\begin{equation}\label{pc}
f_{ij}^{cec} = \int_{w_{l}}^{\infty} w \sqrt{w^{2}-1} (w_{m}+w)^{2}
F(+ Z,w) G_- dw,
\end{equation}
for continuum electron capture (we use natural units in this
section, $\hbar=m_{e}=c=1$).

In Eqs.~\ref{ps} and ~\ref{pc}, $w$ is the total kinetic energy of
the electron including its rest mass, $w_{l}$ is the total capture
threshold energy (rest + kinetic) for electron capture and $F(\pm
Z,w)$ are the Fermi functions. One should note that if the
corresponding positron emission total energy, $w_{m}$, is greater
than -1, then $w_{l}=1$, and if it is less than or equal to 1, then
$w_{l}=\mid w_{m} \mid$. $w_{m}$ is the total $\beta$-decay energy,

\begin{equation}
w_{m} = m_{p}-m_{d}+E_{i}-E_{j},
\end{equation}
where m$_{p}$ and E$_{i}$ are mass and excitation energies of the
parent nucleus, and m$_{d}$ and E$_{j}$ of the daughter nucleus,
respectively.

The number density of electrons associated with protons and nuclei
is $\rho Y_{e} N_{A}$, where $\rho$ is the baryon density, $Y_{e}$
is the ratio of electron number to the baryon number, and $N_{A}$ is
the Avogadro number

\begin{equation}\label{ye}
\rho Y_{e} = \frac{1}{\pi^{2}N_{A}}(\frac {m_{e}c}{\hbar})^{3}
\int_{0}^{\infty} (G_{-}-G_{+}) p^{2}dp,
\end{equation}
where $p=(w^{2}-1)^{1/2}$ is the electron or positron momentum, and
Eq.~\ref{ye} has the units of \textit{moles $cm^{-3}$}. This
equation is used for an iterative calculation of Fermi energies for
selected values of $\rho Y_{e}$ and $T$.

The total continuum electron capture/positron decay rate per unit
time per nucleus is finally given by
\begin{equation}
\lambda^{pd(cec)} =\sum _{ij}P_{i} \lambda _{ij}^{pd(cec)}.
\label{total rate}
\end{equation}
The summation over the initial and final states was carried out
until satisfactory convergence was achieved is stellar weak rates
calculation. Here $P_{i} $ is the probability of occupation of
parent excited states and follows the normal Boltzmann distribution.

It was assumed in our calculation that all daughter excited states,
with energy greater than the separation energy of protons ($S_{p}$)
decay by emission of protons. The proton energy rate from the
daughter nucleus was calculated using
\begin{equation}\label{ln}
\lambda^{p} = \sum_{ij}P_{i}\lambda_{ij}(E_{j}-S_{p}),
\end{equation}
for all $E_{j} > S_{p}$.

The probability of $\beta$-delayed proton emission was calculated by
\begin{equation}\label{pn}
P^{p} =
\frac{\sum_{ij\prime}P_{i}\lambda_{ij\prime}}{\sum_{ij}P_{i}\lambda_{ij}},
\end{equation}
where $j\prime$ are states in the daughter nucleus for which
$E_{j\prime} > S_{p}$. In Eqs.~(\ref{ln}, \ref{pn}),
$\lambda_{ij(\prime)}$ is the sum of the continuum electron capture
and positron decay rates, for the transition $i$ $\rightarrow$
$j(j\prime)$.

\section{\label{sec:level3} Results and discussions}

The \mbox{IBM-1} Hamiltonian given in Eq. (\ref{e_s.ham}) was used
for the description of the structure of the even-even WP
$^{80}$Zr, $^{84}$Mo, $^{88}$Ru, $^{92}$Pd and $^{96}$Cd
nuclei located along the $N$ = $Z$ line in the $A = 80-100$ region.
For the parametrization, similar procedure was performed as
our previous work ~\citep{Nab16}; firstly, the free parameters
were fitted for the known $^{80}$Zr nucleus and then expanded
up to the unknown $^{96}$Cd. For the known $^{80}$Zr, $^{84}$Mo,
$^{88}$Ru and $^{92}$Pd nuclei, the $\epsilon$, ${a_2}$ parameters
given in Hamiltonian (Eq.(\ref{e_s.ham})) and $\overline{\chi}$
in the quadrupole operator (Eq.(\ref{e_term})) were fitted by
using the experimental data from~\cite{NNDC} while those for
the prediction of the unknown $^{96}$Cd nucleus were fitted by using
the results of the shell model calculation given in Ref.~\citep{Coraggio12}.
The appropriate $\epsilon$, $a_{2}$ and $\overline{\chi}$ parameters were
 provided, respectively, by minimizing the root-mean-square ($rms$)
deviation. The best fitted parameters for each nucleus are given in Table~\ref{ta2}.

As seen in Fig.~\ref{f_en1}, the calculated results are in good
agreement with the experimental values for ground state bands. The
energy ratio $R_{4/2}$ in the ground state band is illustrated in
Fig.~\ref{f_en2} for the studied nuclei as a function of boson
numbers $N=10$~$\rightarrow$~$2$. The energy ratio $R_{4/2}=E(4^{+})/E(2^{+})$ gives us some
information to understand the symmetrical characters of the nuclei
and also their shapes.  The positions of each symmetry are also
labeled in Fig.~\ref{f_en2} along with the energy ratio for WP
nuclei. As seen in this figure, $^{80}$Zr is deformed and located in
between $O(6)$-$SU(3)$, $^{84}$Mo is close to $O(6)$, $^{88}$Ru is
in between $U(5)$-$O(6)$ while $^{92}$Pd and $^{96}$Cd are in close
proximity to $U(5)$.

In Fig.~\ref{f_pes1}, the PES of the WP nuclei are plotted as
function of~~$\beta$ and $\gamma$~~in counter plot and also as
function of~~$\beta$~~for~~$\gamma=0$. The fitted
Hamiltonian~(Eq.~\ref{e_s.ham}) parameters, given in
Table~\ref{ta2}, have been used as constants in V($\beta,\gamma$) in
Eq.~\ref{e_s.pes}. According to the plots shown in
Fig.~\ref{f_pes1}, $^{96}$Cd, $^{92}$Pd nuclei are spherical,
$^{88}$Ru is in between spherical and deformed and $^{84}$Mo,
$^{80}$Zr are axially deformed with prolate shape. The energy
surfaces as function of~$\beta$ ($\gamma=0$) are also shown
collectively in Fig.~\ref{f_pes2} including number of bosons ($N$).
Fig.~\ref{f_pes2} shows that $^{96}$Cd nucleus (with $two$ bosons)
and its neighboring nucleus $^{92}$Pd are spherical since their
energy minima are at $\beta=0$. $^{88}$Ru nucleus looks spherical
but its minimum is rather flat so it is not pure spherical. The
energy ratio of $^{88}$Ru nucleus is $R_{4/2}^{EXP}=2.30$ and
$R_{4/2}^{IBM}=2.29$ (see Fig.~\ref{f_en2}) and located in between
$U(5)$--$O(6)$ like the critical point
E(5)~\citep{Iachello00i,Casten00} since its corresponding energy
ratio is $R_{4/2}^{E(5)}=2.19$. The energy surface of $^{88}$Ru is
rather flat similar to the energy surface of E(5) shown in Fig.~2 of
Ref.~\citep{Iachello00i} given for the $U(5)$--$O(6)$ shape phase
transitions. $^{84}$Mo is also deformed and looks like prolate shape
($\beta_{min} > 0$), but its energy minima (of both oblate and
prolate sides) are close to each other and its minimum in the
counter plot is quite wider. As seen from the energy ratio $R_{4/2}$
given in Fig.~\ref{f_en2}, $^{84}$Mo nucleus is close to $O(6)$.
$^{80}$Zr with $10$ bosons was also deformed with prolate shape. The
minimum on oblate side is not so high as compared with prolate side
and this nucleus can be located in between $O(6)$-$SU(3)$. The
$\beta_{min}$ of all WP nuclei were calculated and given in
Table~\ref{ta1}.

The \mbox{pn-QRPA} calculated  GT$^{+}$ strength distributions for
$^{80}$Zr, $^{84}$Mo and $^{88}$Ru are presented in upper panels of
Fig.~\ref{80-88}. Lower panel shows the strength distributions
calculated by Sarriguren \citep{Sar11}. Sarriguren used the Skyrme
Hartree-Fock + BCS + QRPA model for his calculation of strength
distributions. Details of the formalism of Sarriguren's calculation
may be seen in \citep{Sar09a,Sar09b}. It is to be noted that
Sarriguren used a quenching factor of 0.55 in his calculation. This
may be compared with quenching factor of 0.6 used in our
calculation. It may be noted, at the outset, that our calculated
strength distribution is more fragmented. The peak strengths in the
daughter 1$^{+}$ states match relatively well for the case of
$^{80}$Zr. Few of the low-lying strengths were not calculated by
Sarriguren for the case of $^{84}$Mo. Similarly it can be seen that
centroid of our calculated GT$^{+}$ strength distribution for the
case of $^{88}$Ru resides at a much lower energy in daughter. The
peak strength calculated by Sarriguren in this nucleus is calculated
around 8 MeV in daughter. Our \mbox{pn-QRPA} calculated strength
distributions of the remaining two $N$ = $Z$ WP nuclei, $^{92}$Pd
and $^{96}$Cd, along with comparison of Sarriguren's calculation is
shown in Fig.~\ref{92-96}. Whereas bulk of strength was calculated
in a single transition by Sarriguren (see the transition at around 7
MeV in daughter $^{96}$Ag), we notice that the strength is rather
fragmented into multiple closely lying daughter states in our model.
The fragmentation of GT strength is a nice feature present in our
model and has consequences for the calculation of weak rates which
we present later in this section. The calculated total GT strength
matches well for the two calculations only in the case of $^{92}$Pd.
For remaining cases we calculate a much bigger total strength
specially for the case of $^{80}$Zr and $^{84}$Mo. Correspondingly
our calculated electron capture rates for these two nuclei are
roughly factor two bigger for $rp$-process temperatures and would be
discussed later.

Fig.~\ref{hl} compares our calculated terrestrial half-lives for
these neutron-deficient heavy WP nuclei with experimental and
previous calculations. Experimental half-lives were taken from the
recent atomic mass evaluation AME2012 \citep{Aud12}. Shown also in
Fig.~\ref{hl} are the Hartree-Fock (HF) and QRPA calculations using
the Sk3 \citep{Bei75} and SG2 \citep{Gia81} forces performed by
Sarriguren and collaborators \citep{Sar05}. Biehle and Vogel
\citep{Bie92} also performed a QRPA calculation, which is in very
good agreement with measured half-life for $^{88}$Ru and $^{92}$Pd,
and is also shown in Fig.~\ref{hl}. None of the previous
calculations reported half-life calculation for $^{96}$Cd. It is
noted that HF half-lives are systematically lower than the
corresponding QRPA and experimental values. It is well known fact
that the QRPA correlations tend to reduce the mean-field
Gamow-Teller strength thereby increasing the calculated half-life
values. Our calculated half-lives are in good agreement with
experimental values. Our calculated percentage deviation from
measured values for $^{80}$Zr, $^{84}$Mo, $^{88}$Ru, $^{92}$Pd and
$^{96}$Cd are 21$\%$, 71$\%$, 17$\%$, 2$\%$ and 17$\%$,
respectively. Fig.~\ref{def} shows how the \mbox{pn-QRPA} calculated
half-lives vary as a function of the deformation parameter. Here the
dotted lines show the measured half-lives. It can be seen from
Fig.~\ref{def} that the deformations from \citep{Lal99} reproduce
well the measured half-lives of these $N = Z$ nuclei.

We calculated the electron capture cross sections (ECC) for typical
collapse temperatures ($T \sim 0.5 - 2.0$ MeV). At these finite
temperatures, the GT transitions are unblocked as the residual
interaction becomes sufficiently strong. From astrophysical (as well
as experimental) viewpoint, the important range of the incident
electron energy ($E_{e}$) is up to 30 MeV \citep{Gia15}. Accordingly
we present our calculated ECC in Fig.~\ref{cs} up to $E_{e}$ = 30
MeV. Fig.~\ref{cs} shows our calculated ECC for $^{80}$Zr,
$^{84}$Mo, $^{88}$Ru, $^{92}$Pd and $^{96}$Cd as a function of
$E_{e}$.  The $Q$ value of the reaction signifies the minimum
electron energy to initiate the electron capture process (at finite
temperatures, this threshold is further lowered by the internal
excitation energy). The ECC increases drastically within the first
couple of MeV of $E_{e}$ above threshold and maybe traced to the
behavior of calculated GT distribution.  At low incident electron
energies the capture process is sensitive to the details of
calculated finite temperature GT strength distribution.  Because of
the $(E_e-Q+E_i-E_f)^{2}$ factor in Eq.~(14),  the calculated cross
sections display a more gradual increase for electron energy $E_{e}
\ge$ 10 MeV.   With increasing $E_{e}$ the ECC continues to rise
modestly. We witness a notable increase up to two orders of
magnitude in the calculated ECC as the temperature increases from
0.5 MeV to 1.0 MeV. This corresponds to a significant thermal
unblocking of the GT$^{+}$ channel (the Fermi contribution to the
total ECC is negligible). Because majority of the transitions are
already unblocked at T = 1.0 MeV, a further increase in temperature
to T = 1.5 MeV and 2.0 MeV, results in roughly one order of
magnitude enhancement of the calculated ECC. A more or less similar
temperature dependence of calculated ECC is witnessed for $^{84}$Mo,
$^{88}$Ru, $^{92}$Pd and $^{96}$Cd and shown in Fig.~\ref{cs}.

Figs.~\ref{80Zrph} to~\ref{96Cdph} depict the \mbox{pn-QRPA}
calculated stellar rates of $\beta$-delayed proton emission (top
panel) and the $\beta$-delayed proton emission probabilities
($P$$_{p}$) (middle panel) for the $N = Z$ heavy waiting point
nuclei. The rates are shown as a function of core temperature and
density. The bottom panel shows the corresponding calculation of
phase space for electron capture as given in Eq.~\ref{pc}.
Fig.~\ref{80Zrph} shows the calculated weak rates and phase space
factors for $^{80}$Zr. The phase space factors decrease slightly as
temperature soars from 0.5--2.0 GK (at density $\rho$ =
10$^{5}$-10$^{6}$ gm.cm$^{-3}$). As temperature and density increase
further, the phase space increases monotonically. The proton
emission rates are given in units of $MeV.s^{-1}$. The rates are
calculated for density and temperature range believed to be relevant
for the $rp$-process. The proton emission rates decrease slightly as
temperature soars from 0.5--1.0 GK at density $\rho$ = 10$^{5}$
gm.cm$^{-3}$. Else the rates increase with increasing core
temperature and density. The initial decrease in rates can be traced
to reduction in phase space.  The calculated rates follow a similar
trend as that of the corresponding phase space factors.
Fig.~\ref{84Moph} shows the calculated rates and phase space factors
for $^{84}$Mo. Here one notes a slight reduction in phase space at
density $\rho$ = 10$^{5}$-10$^{6}$ gm.cm$^{-3}$ as temperature
increases from 1.0--1.5 GK. The proton emission rates follow a more
or less similar trend. Fig.~\ref{88Ruph}, Fig.~\ref{92Pdph} and
Fig.~\ref{96Cdph} show phase space and rate calculations for
$^{88}$Ru, $^{92}$Pd and $^{96}$Cd, respectively, and follow a
similar trend as Fig.~\ref{80Zrph}.

The $rp$-process weak-interaction mediated rates of heavy WP nuclei
were calculated using the \mbox{pn-QRPA} model. Figs.~\ref{80Zr}
to~\ref{96Cd} show the calculated weak-interaction mediated rates
for the WP nuclei as a function of stellar temperature and density.
The left-panels of these figures show the stellar electron capture
(cEC) and positron decay ($\beta^{+}$) rates as a function of the
stellar temperature for selected density of 10$^{5}$, 10$^{6}$,
10$^{6.5}$ and 10$^{7}$ g.cm$^{-3}$ (pertinent to $rp$-process
conditions). The positron decay rates remain constant as the stellar
density increases by two orders of the magnitude. The right panels
show the total sum of these two rates. The upper panels are the
Skyrme HF+BCS+QRPA calculation of Sarriguren and reproduced from
\cite{Sar11} whereas the lower panels depict our calculation. It is
to be noted that Sarriguren calculated his weak rates only up to
stellar temperature of 10 GK. We performed our calculation up to a
higher temperature of 30 GK.  The temperature range considered to be
most relevant with the $rp$-process is shown shaded in the figures.
Calculated weak rates for $^{80}$Zr are depicted in Fig.~\ref{80Zr}.
It can be noted that our positron decay rates are around a factor 2
bigger than Sarriguren rates under $rp$-process conditions. At
temperatures of 10 GK our positron decay rates are factor 20 bigger.
Our calculated electron capture rates are also factor 2 bigger than
Sarriguren rates for $rp$-process conditions. Our calculated total
rates are factor 2 bigger for $rp$-process conditions as compared to
Sarriguren calculated total rates. Another feature to be noted is
that for $rp$-process conditions, our calculated positron decay
rates are factor (6~--~20) bigger at $\rho$ = 10$^{5}$ gm.cm$^{-3}$
when compared with our calculated electron capture rates. However as
stellar density and temperature increases so do the electron capture
rates and at density $\rho$ = 10$^{7}$ gm.cm$^{-3}$, the calculated
electron capture rates increases and are factor 4 bigger than the
competing positron decay rates.

For the case of $^{84}$Mo, the Skyrme HF+BCS+QRPA calculated
positron decay rates are roughly factor 1.5 bigger than
\mbox{pn-QRPA} rates (Fig.~\ref{84Mo}) for $rp$-process conditions.
However our electron capture rates are around a factor two bigger
and the total rates are in overall good agreement with Sarriguren's
rates. The HF+BCS+QRPA calculated positron decay rates for $^{88}$Ru
are roughly twice our rates whereas the electron capture rates are
in good comparison under $rp$-process conditions (see
Fig.~\ref{88Ru}). Consequently the total rate comparisons between
the two calculations improve as density stiffens from $\rho$ =
10$^{4}$ gm.cm$^{-3}$ to $\rho$ = 10$^{7}$ gm.cm$^{-3}$.
Fig.~\ref{92Pd} and Fig.~\ref{96Cd} show excellent agreement between
\mbox{pn-QRPA} and Skyrme HF+BCS+QRPA calculated rates for $^{92}$Pd
and $^{96}$Cd, respectively, under $rp$-process conditions. For all
nuclei our calculated rates are enhanced, up to a factor 20, at high
temperatures. Convergence of the rate calculation is an important
consideration as temperature increases to 10 GK and beyond (see Eq.
(\ref{total rate})). This is primarily due to finite occupation
probability of parent excited states. We took 200 initial and 300
final states in our rate calculation which ensured satisfactory
convergence in our rate calculation. We further note that
convergence was achieved in our rate calculations for excitation
energies well in excess of 10 MeV due to the availability of a huge
model space (up to 7 major oscillator shells) in our \mbox{pn-QRPA}
model. The self-consistent approach of the Skyrme HF+BCS+QRPA
calculation forces one to use limited configuration spaces and might
lead to convergence problem in rate calculation at high stellar
temperature. We also note that at density $\rho$ = 10$^{7}$
gm.cm$^{-3}$ and for $rp$-process temperature range, electron
capture rates are bigger, up to a factor 5, than the competing
positron decay rates. Table~\ref{ta4} shows the ratio of the
\mbox{pn-QRPA} calculated electron capture and positron decay rates,
at selected density of 10$^{6}$ g.cm$^{-3}$, as a function of
stellar temperature, for the five $N$~=~$Z$ WP nuclei. It can be
noted in all cases that electron capture rates become significant
with increasing stellar temperatures. This is a curious finding of
this work and might prove useful for  nuclear reaction network
calculations.

\section{\label{sec:level4} Conclusions}

For this work we selected five, $N$~=~$Z$, WP nuclei, namely
$^{80}$Zr, $^{84}$Mo, $^{88}$Ru, $^{92}$Pd and $^{96}$Cd, to study
their nuclear shapes and to calculate their half-lives and
associated stellar weak rates. The plot of the PES, as a function
deformation parameters, is quite useful for the prediction of the
geometric shapes for the given nuclei. According to IBM
calculations, $^{88}$Ru, $^{92}$Pd and $^{96}$Cd are found to be
spherical, whereas $^{80}$Zr and $^{84}$Mo are deformed. Moreover,
the energy ratio of $^{84}$Mo and $^{88}$Ru nuclei are close to the
typical ratio of $O(6)$ and $E(5)$, respectively. $^{84}$Mo nucleus
could be $\gamma$-unstable and $^{88}$Ru is located in between
$U(5)$ and $O(6)$. It is noted that the structure of given WP nuclei
in the $N$~=~$Z$ is rather complex and it is suggested that some
useful and more sophisticated nuclear models may be used to study
their structure and associated shape transitions in detail.

We calculated the GT strength distributions, $\beta$-decay
half-lives, electron capture cross-sections, positron decay $\&$
continuum electron capture rates, energy rates of $\beta$-delayed
protons and their emission probabilities for the heavy WP nuclei.
The calculated half-lives were in good agreement with the
experimental half-lives determined from the recent atomic mass
evaluation AME2012. We noted that a significant thermal unblocking
of the GT$^{+}$ channel occurs as temperature increases from
0.5--1.0 MeV. Our total weak rates are a factor two bigger than the
Skyrme HF+BCS+QRPA calculated rates for $^{84}$Mo. For remaining
$N$~=~$Z$ nuclei the two calculations are in  good agreement under
$rp$-process conditions. Our calculated rates are bigger than the
HF+BCS+QRPA rates at high stellar temperatures. For $rp$-process
conditions, as the density stiffens, the calculated electron capture
rates become more and more significant. At density $\rho$ = 10$^{7}$
gm.cm$^{-3}$, our electron capture rates are bigger, up to a factor
5, than the competing positron decay rates. The study reconfirms the
conclusion made by Sarriguren that electron capture rates form an
integral part of weak-interaction mediated rates under $rp$-process
conditions and should not be neglected in nuclear reaction network
calculations as done in past (e.g. \cite{Sch98}).

\section*{Acknowledgements}

J.-U. Nabi also wishes to acknowledge the support provided by the
Higher Education Commission (Pakistan) through the HEC Project No.
20-3099.

\newpage
\onecolumn

\newpage
\begin{figure}
\includegraphics[width=16cm]{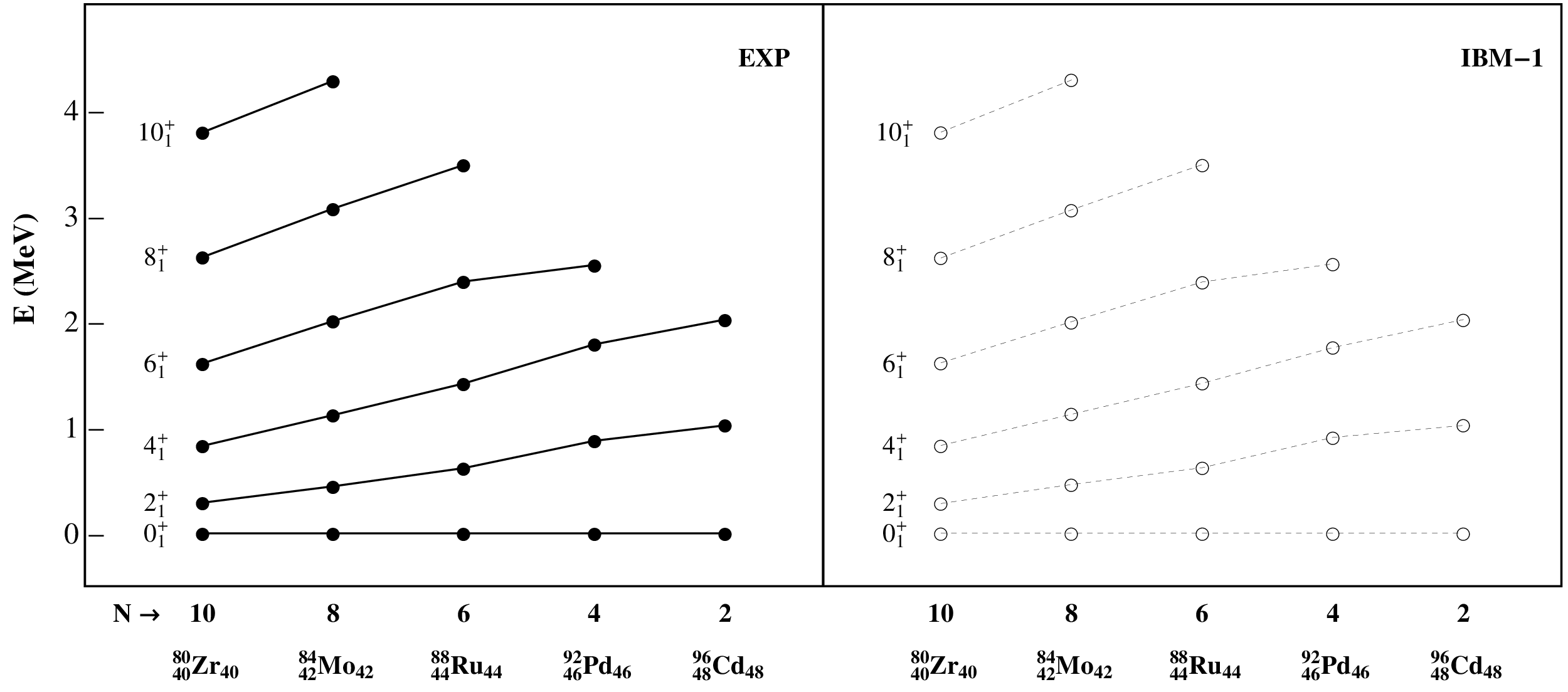}
\caption{The experimental energy spectra (left panel) and
the calculated energy spectra (right panel)
of ground-state bands for $^{80}$Zr, $^{84}$Mo,
$^{88}$Ru, $^{92}$Pd and $^{96}$Cd nuclei.} \label{f_en1}
\end{figure}
\clearpage
\begin{figure}
\includegraphics[width=13cm]{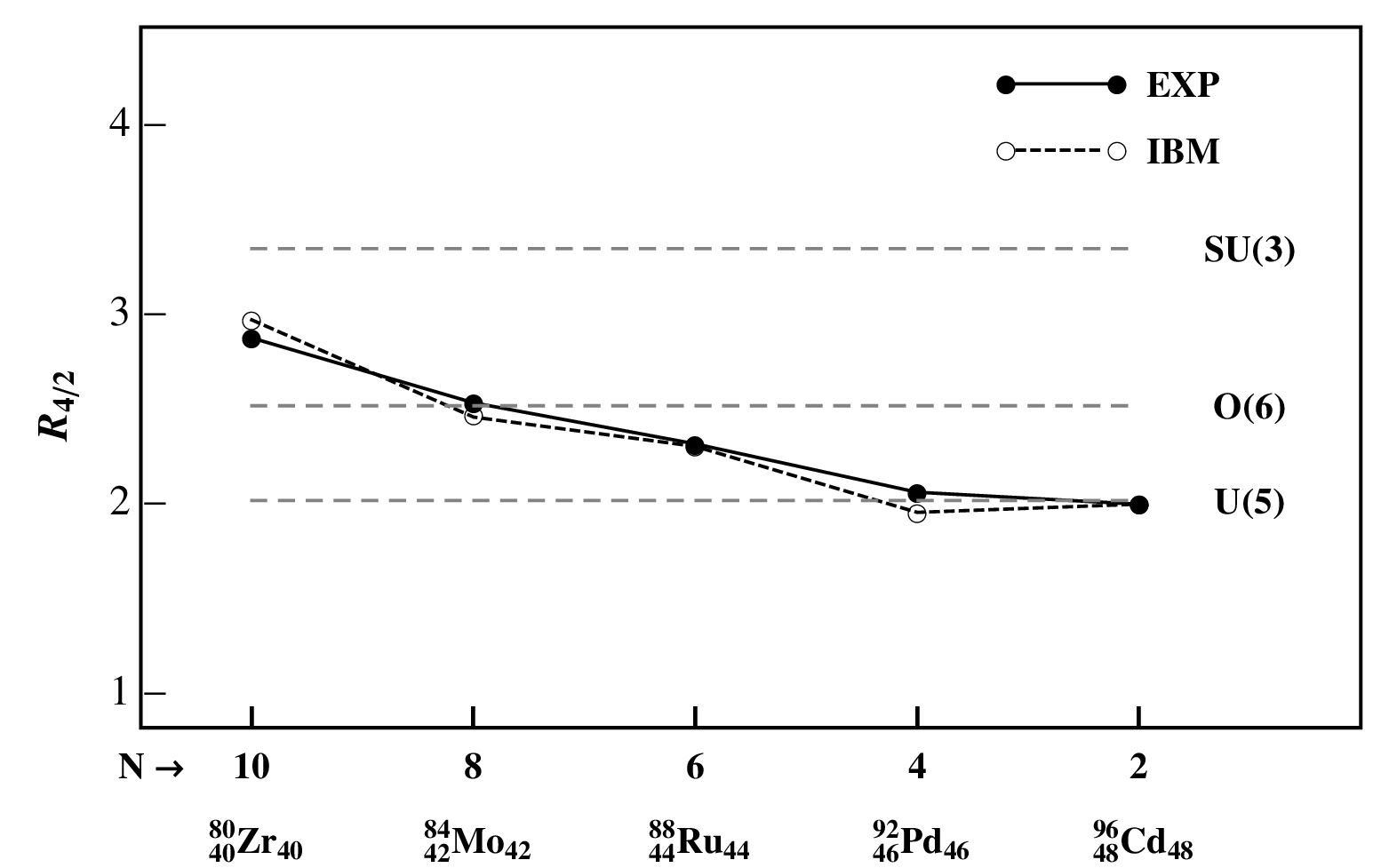}
\caption{Measured and calculated values of the energy ratio
$R_{4/2}$ as function of boson numbers ($N$) for WP nuclei.}
\label{f_en2}
\end{figure}
\clearpage
\begin{figure}
\includegraphics[width=17cm]{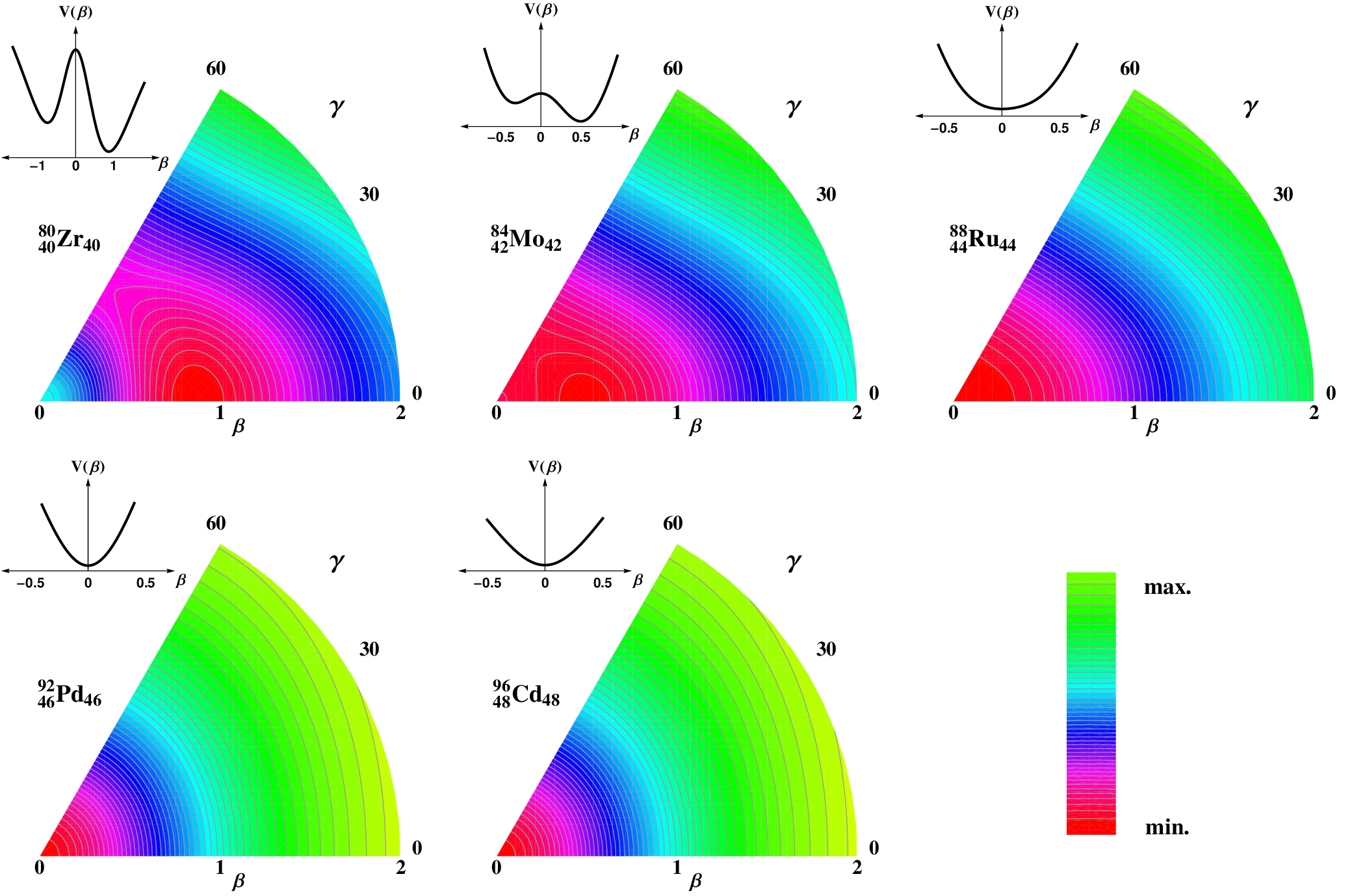}
\caption{(Color online) PES's for WP nuclei. The graphs show the
energy surface as a function of~~$\beta$~~for~~$\gamma=0$.}
\label{f_pes1}
\end{figure}
\clearpage
\begin{figure}
\includegraphics[width=8cm]{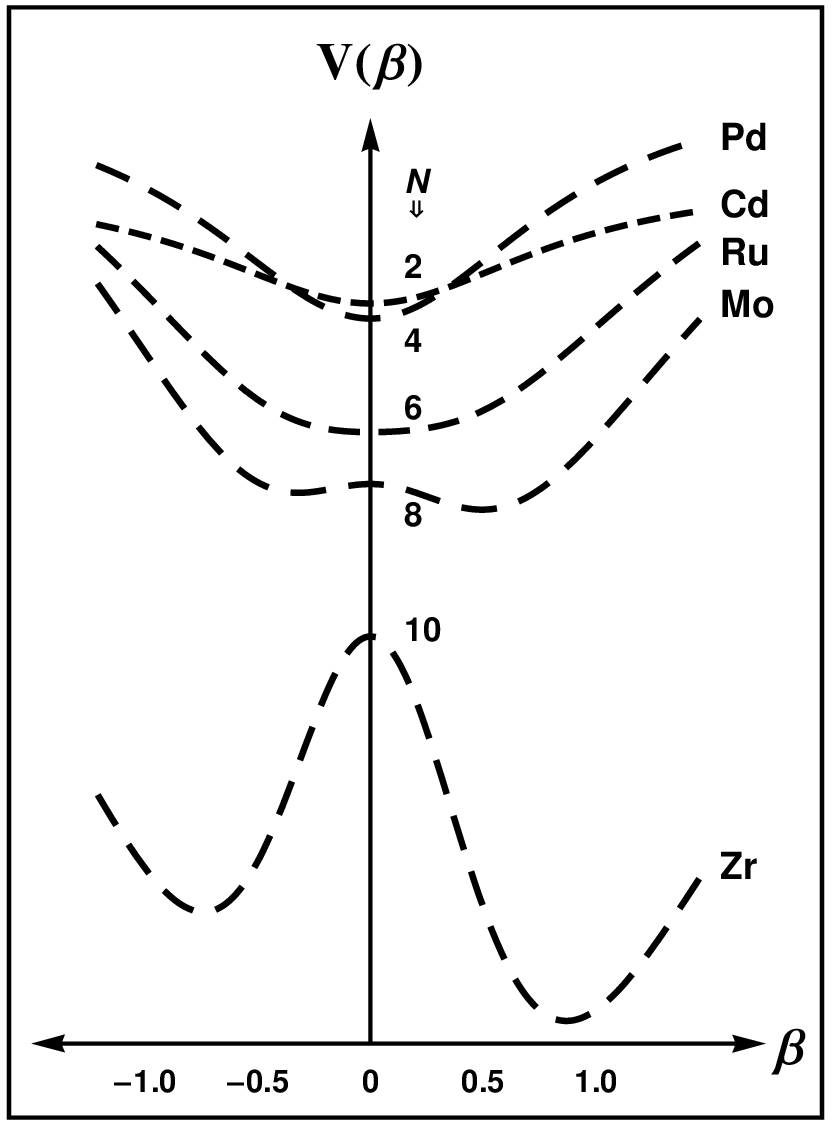}
\caption{The energy surfaces as function of $\beta$ for WP nuclei. The
number of bosons ($N$) is also given in the plot.} \label{f_pes2}
\end{figure}
\clearpage
\begin{figure}
\includegraphics[width=15cm]{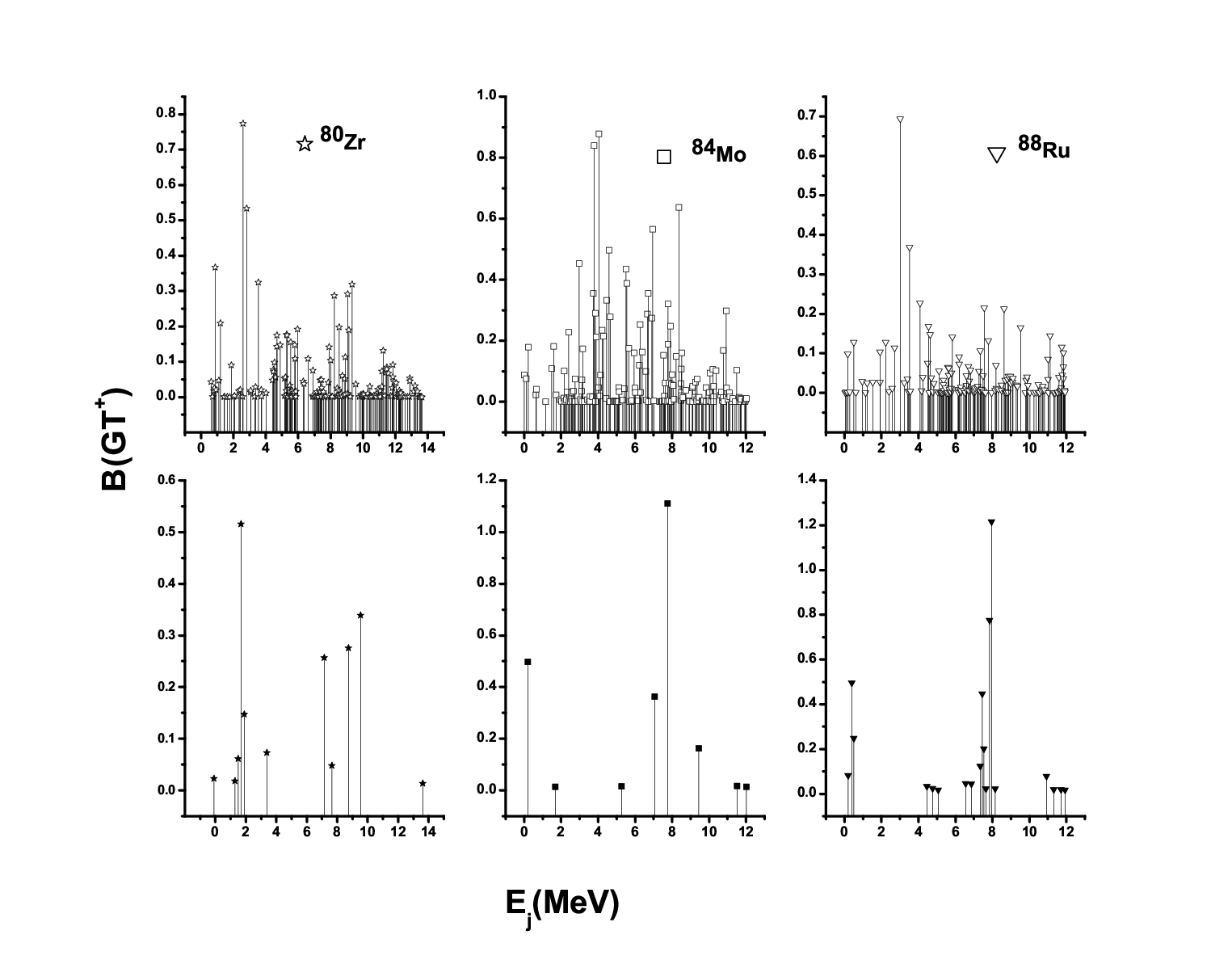}
\caption{Calculated BGT$^{+}$ strength distribution for $^{80}$Zr,
$^{84}$Mo and $^{88}$Ru (upper panels for this work and lower panels for previous work performed
by Sarriguren \cite{Sar11}).}
\label{80-88}
\end{figure}
\clearpage
\begin{figure}
\includegraphics[width=15cm]{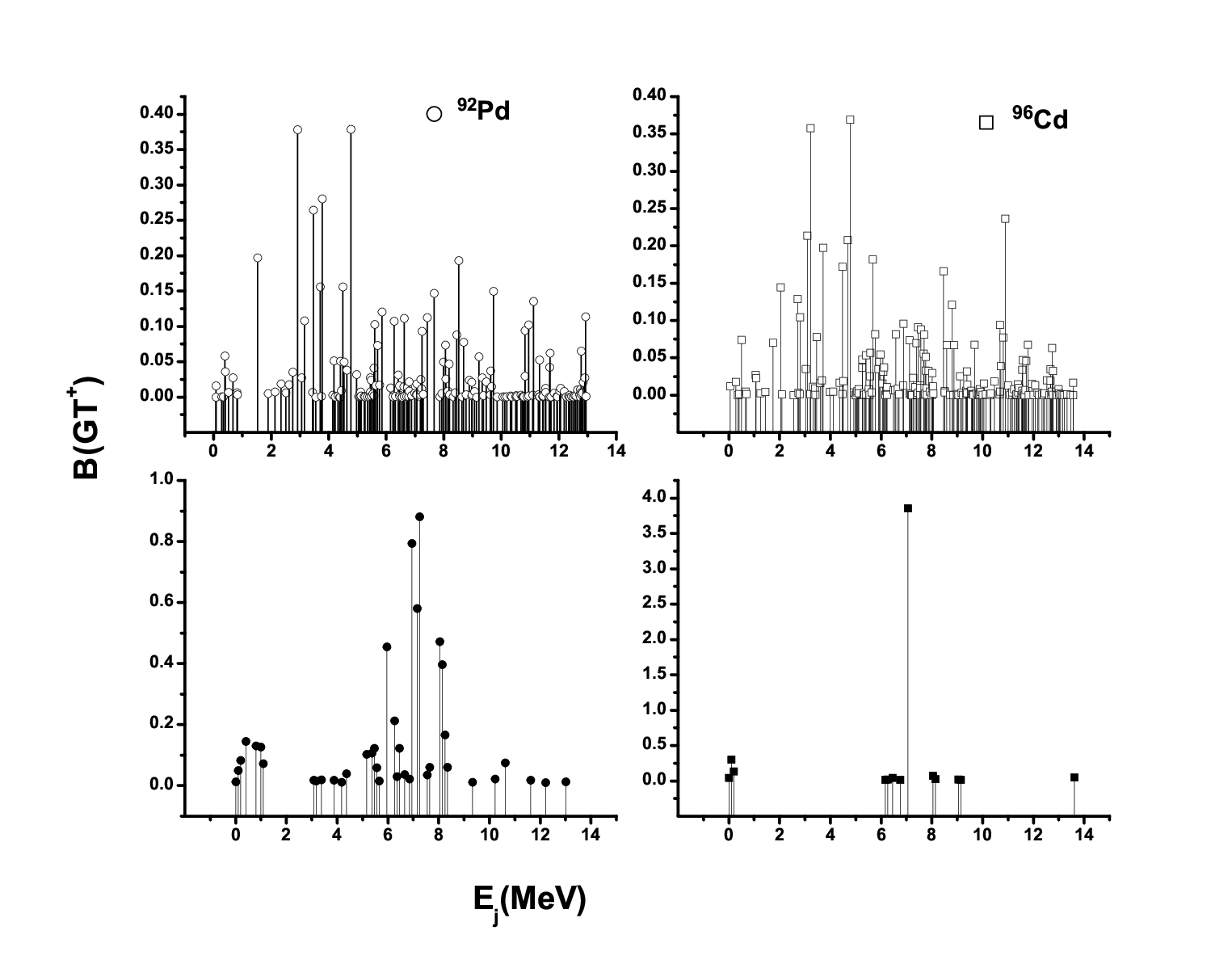}
\caption{Same as Fig.~\ref{80-88} but for $^{92}$Pd and $^{96}$Cd.}
\label{92-96}
\end{figure}
\clearpage
\begin{figure}
\includegraphics[width=5.0in,height=4.3in]{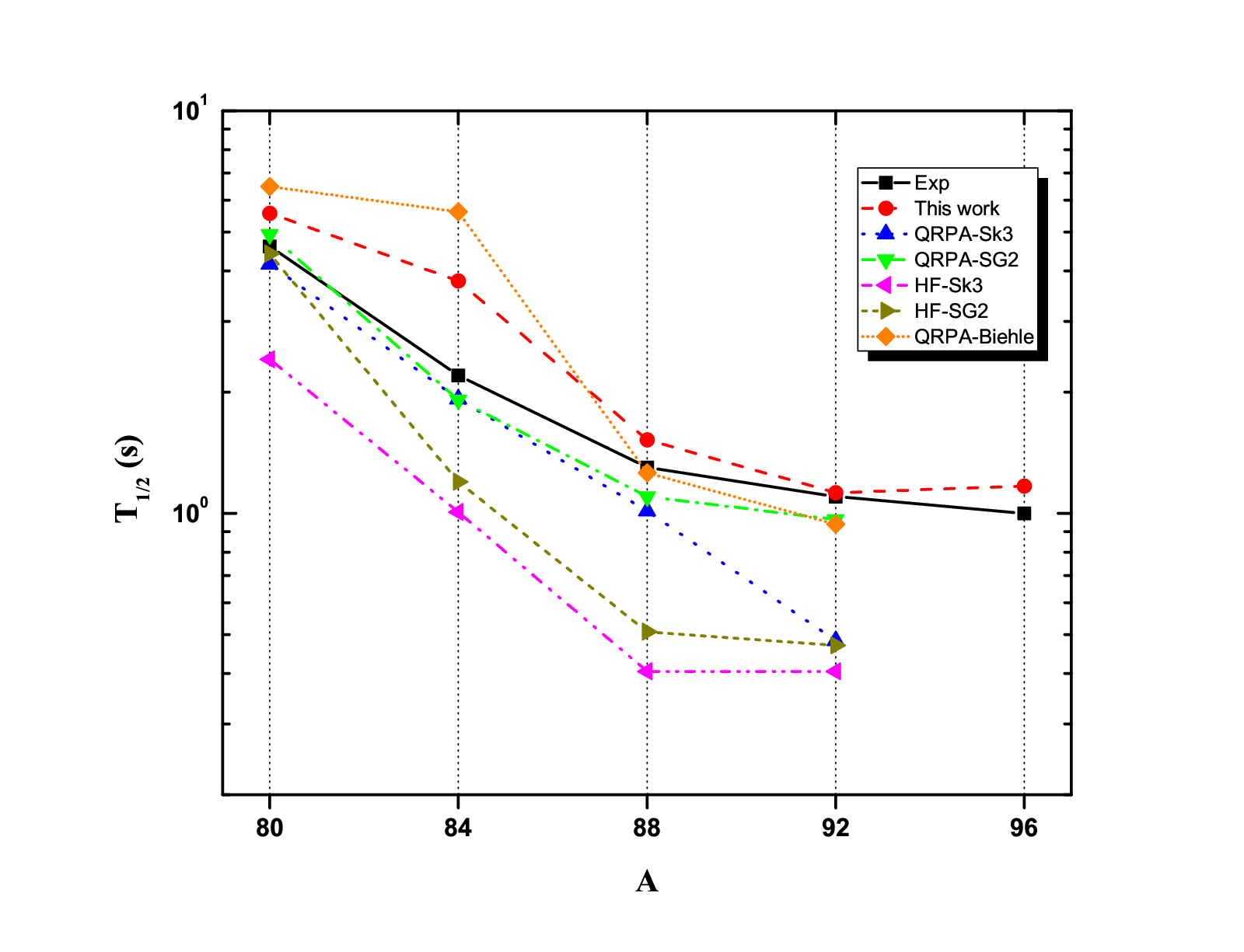}
\caption {(Color online) The comparison of the experimental
half-lives of WP nuclei with present and previous calculations.}
\label{hl}
\end{figure}
\clearpage
\begin{figure}
\includegraphics[width=4.3in,height=4.3in]{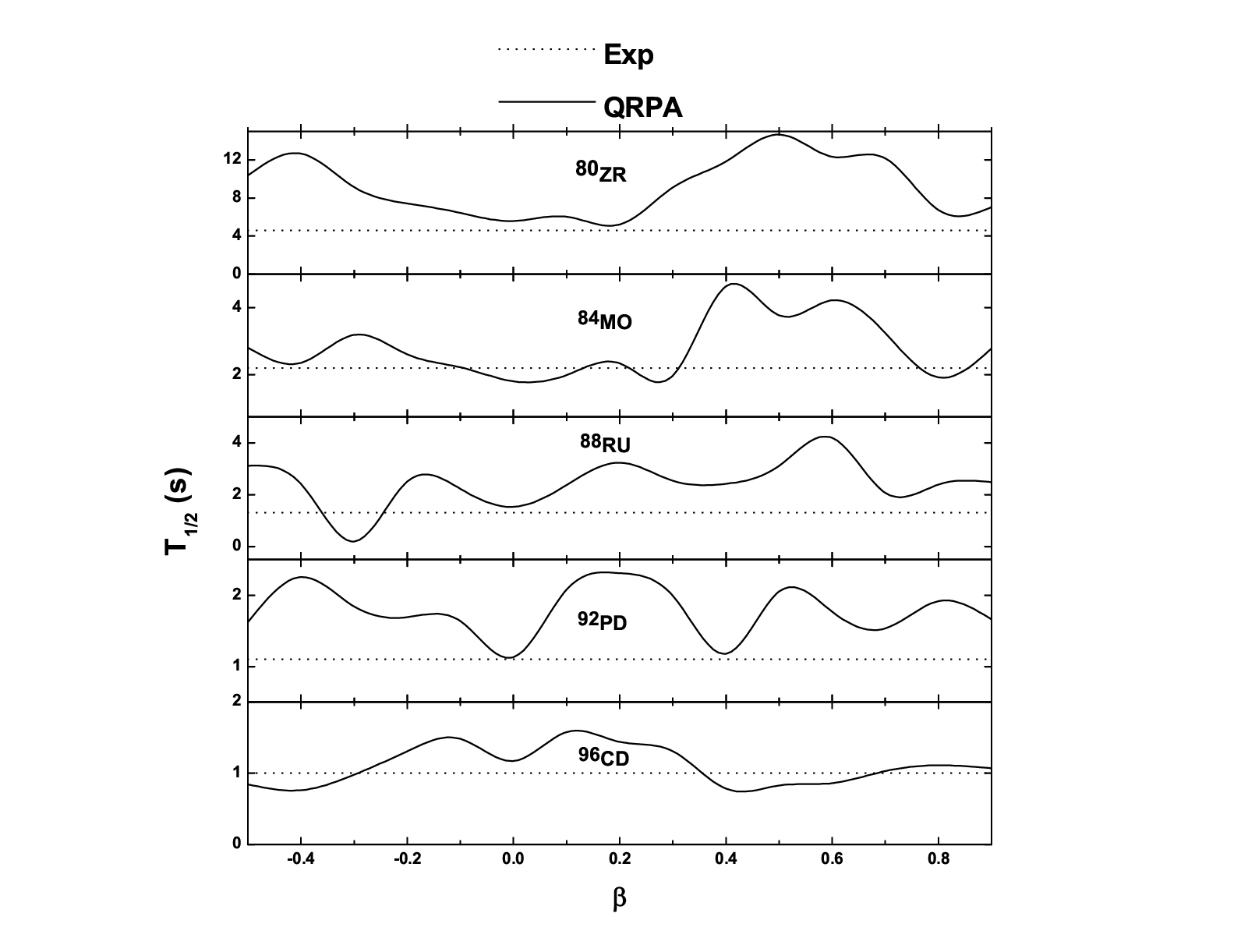}
\caption {Dependence of the calculated half-lives on the values of
deformation parameter for $N$ = $Z$ nuclei. The dotted line shows
the measured half-life.} \label{def}
\end{figure}
\clearpage
\begin{figure}
\includegraphics[width=5.9in,height=4.3in]{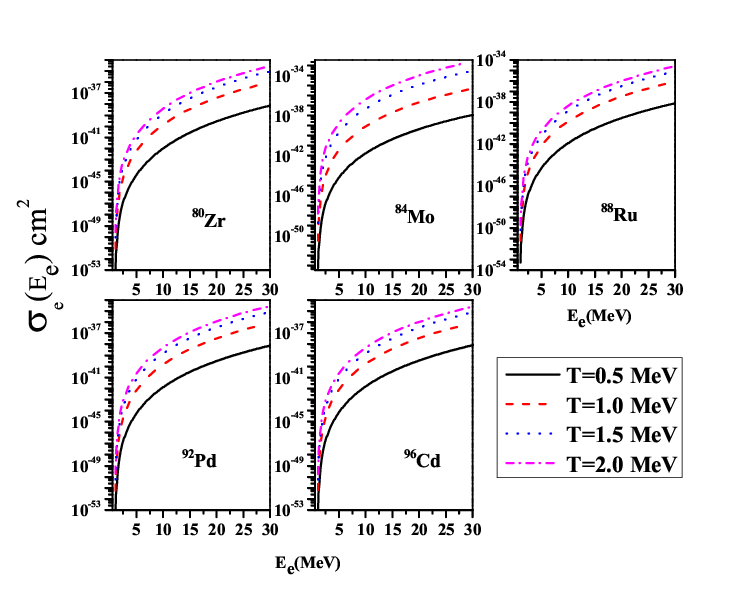}
\caption {(Color online) Electron capture cross sections for $N$ =
$Z$ nuclei, using the \mbox{pn-QRPA} theory, as a function of the
incident electron energy ($E_{e}$) at different stellar
temperatures.}\label{80cs} \label{cs}
\end{figure}
\clearpage
\begin{figure}
\includegraphics[width=4.3in,height=4.3in]{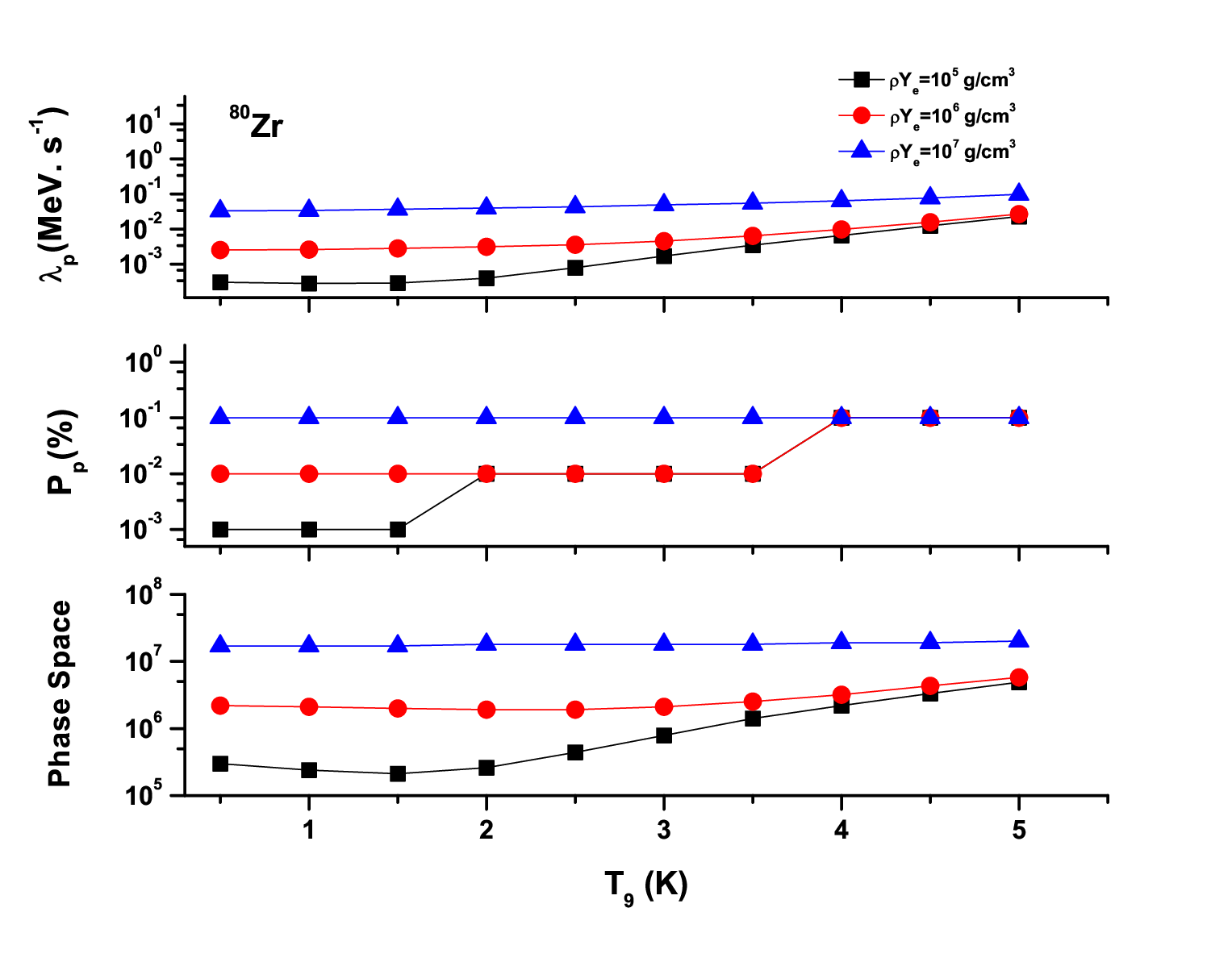}
\caption {(Color online) The \mbox{pn-QRPA} calculated  energy rates
of $\beta$-delayed protons (top panel), probabilities of
$\beta$-delayed proton emissions (middle panel) and phase space
(bottom panel) for $^{80}$Zr as a function of stellar temperature
and density.} \label{80Zrph}
\end{figure}
\clearpage
\begin{figure}
\includegraphics[width=4.3in,height=4.3in]{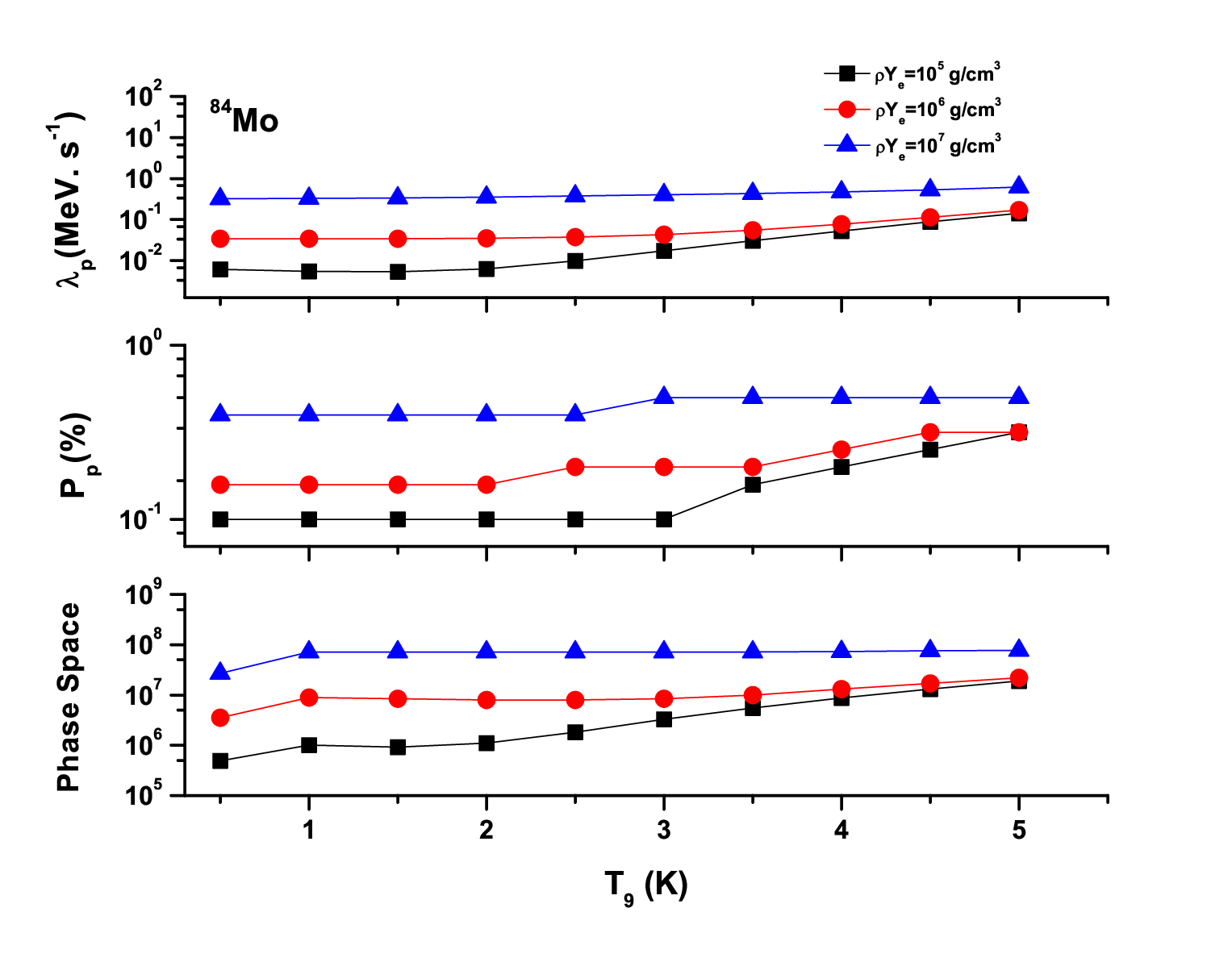}
\caption {(Color online) Same as Fig.~\ref{80Zrph} but for
$^{84}$Mo.} \label{84Moph}
\end{figure}
\clearpage
\begin{figure}
\includegraphics[width=4.3in,height=4.3in]{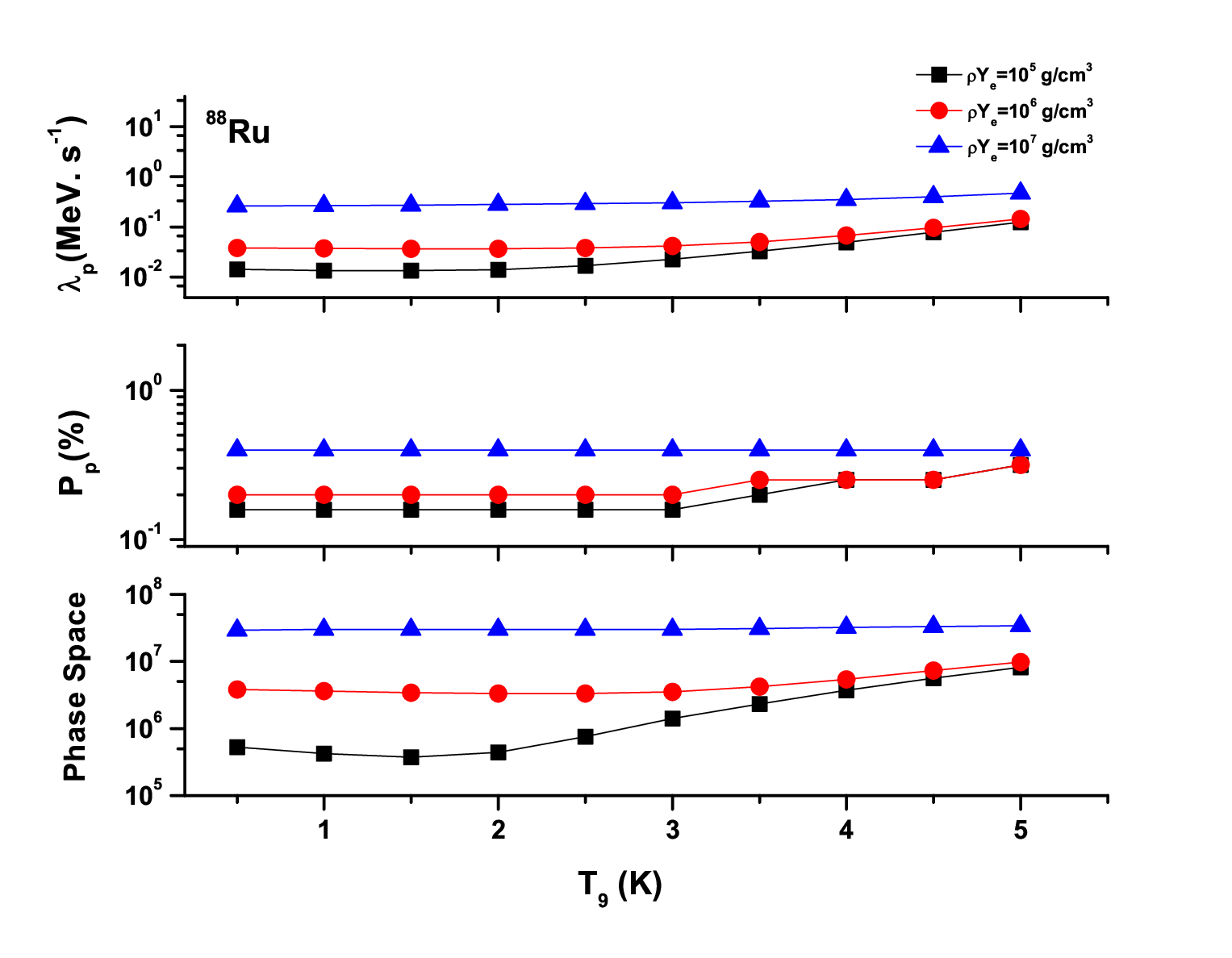}
\caption {(Color online) Same as Fig.~\ref{80Zrph} but for
$^{88}$Ru.} \label{88Ruph}
\end{figure}
\clearpage
\begin{figure}
\includegraphics[width=4.3in,height=4.3in]{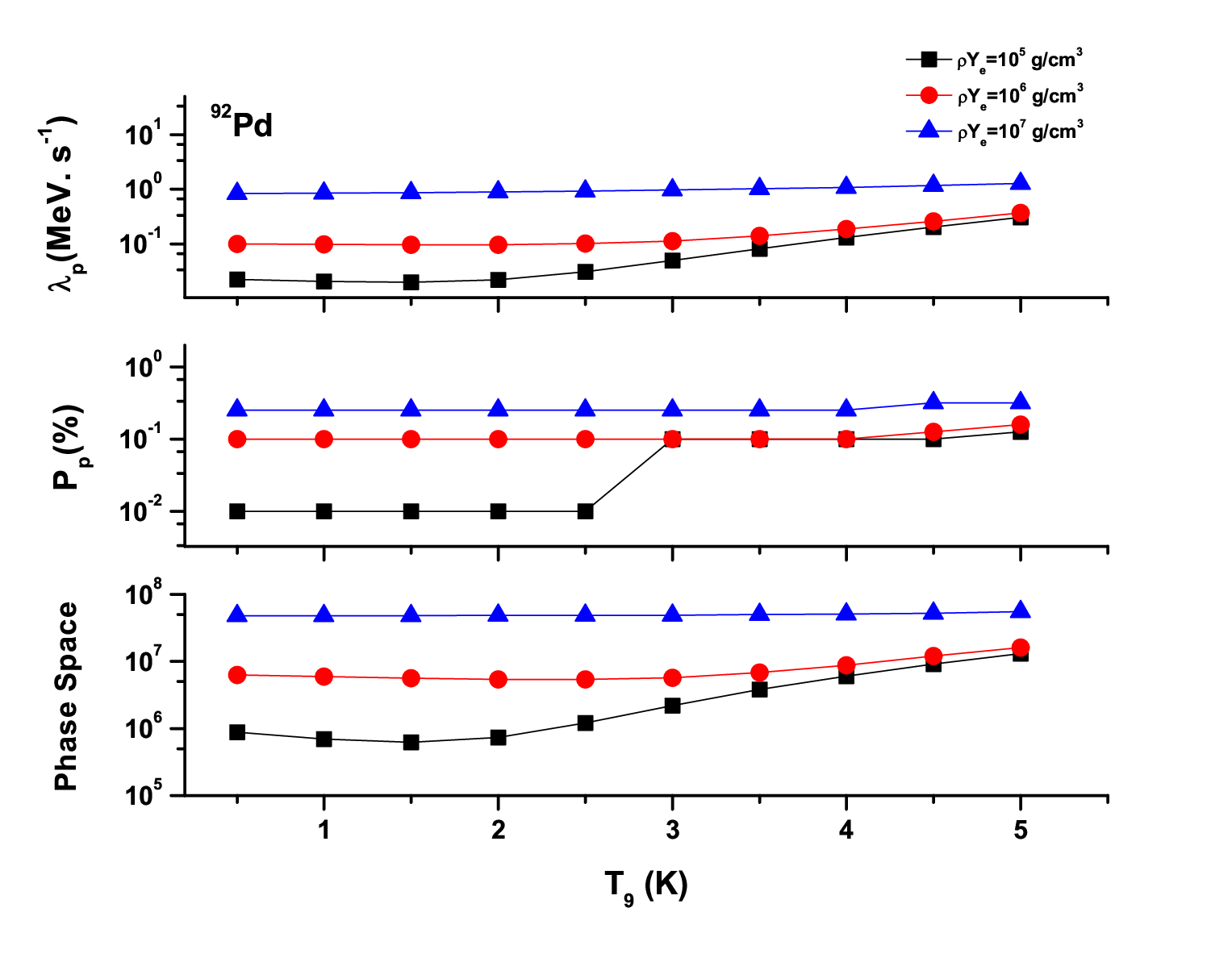}
\caption {(Color online) Same as Fig.~\ref{80Zrph} but for
$^{92}$Pd.} \label{92Pdph}
\end{figure}
\clearpage
\begin{figure}
\includegraphics[width=4.3in,height=4.3in]{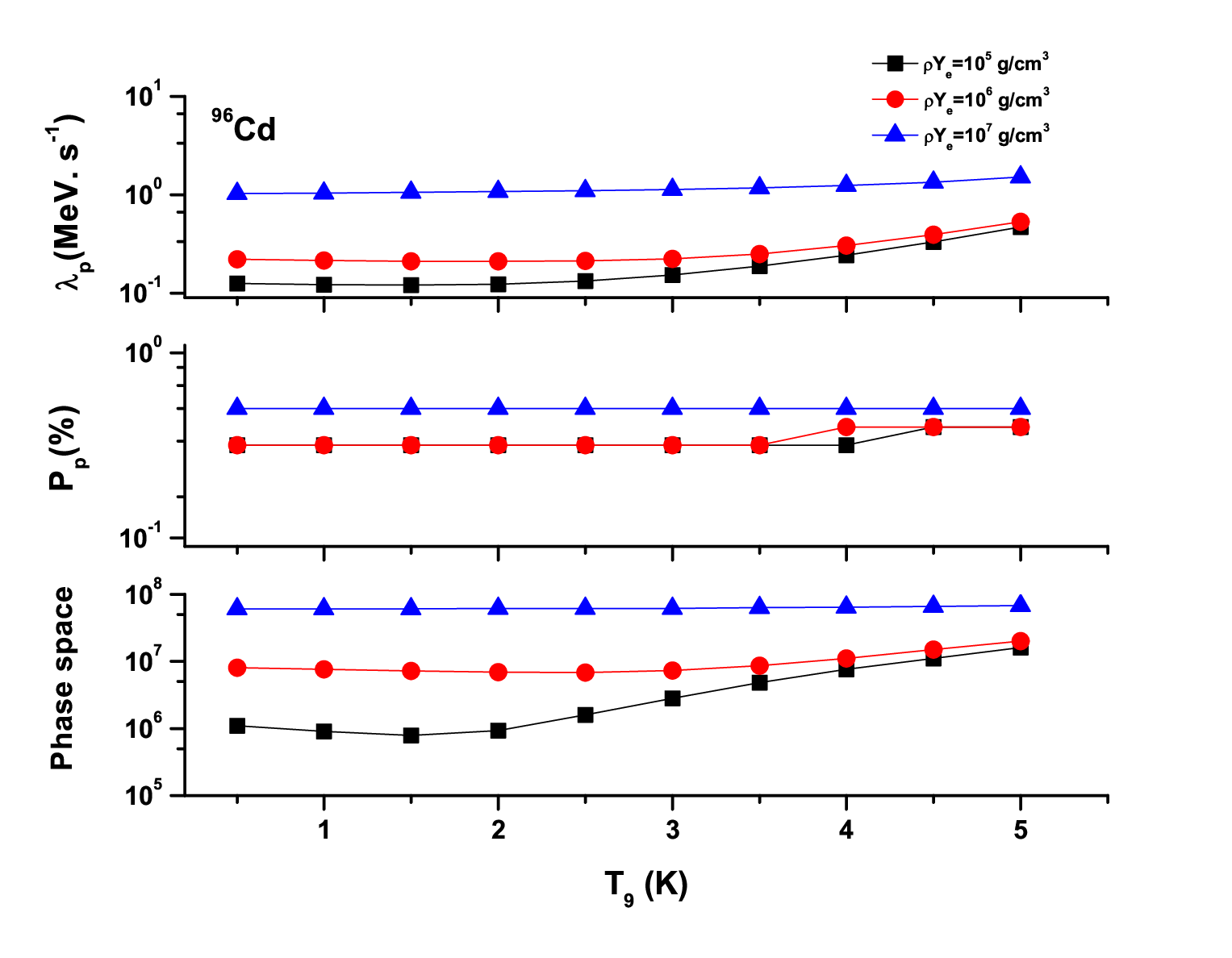}
\caption {(Color online) Same as Fig.~\ref{80Zrph} but for
$^{96}$Cd.} \label{96Cdph}
\end{figure}
\clearpage
\begin{figure}
\includegraphics[width=4.3in,height=4.3in]{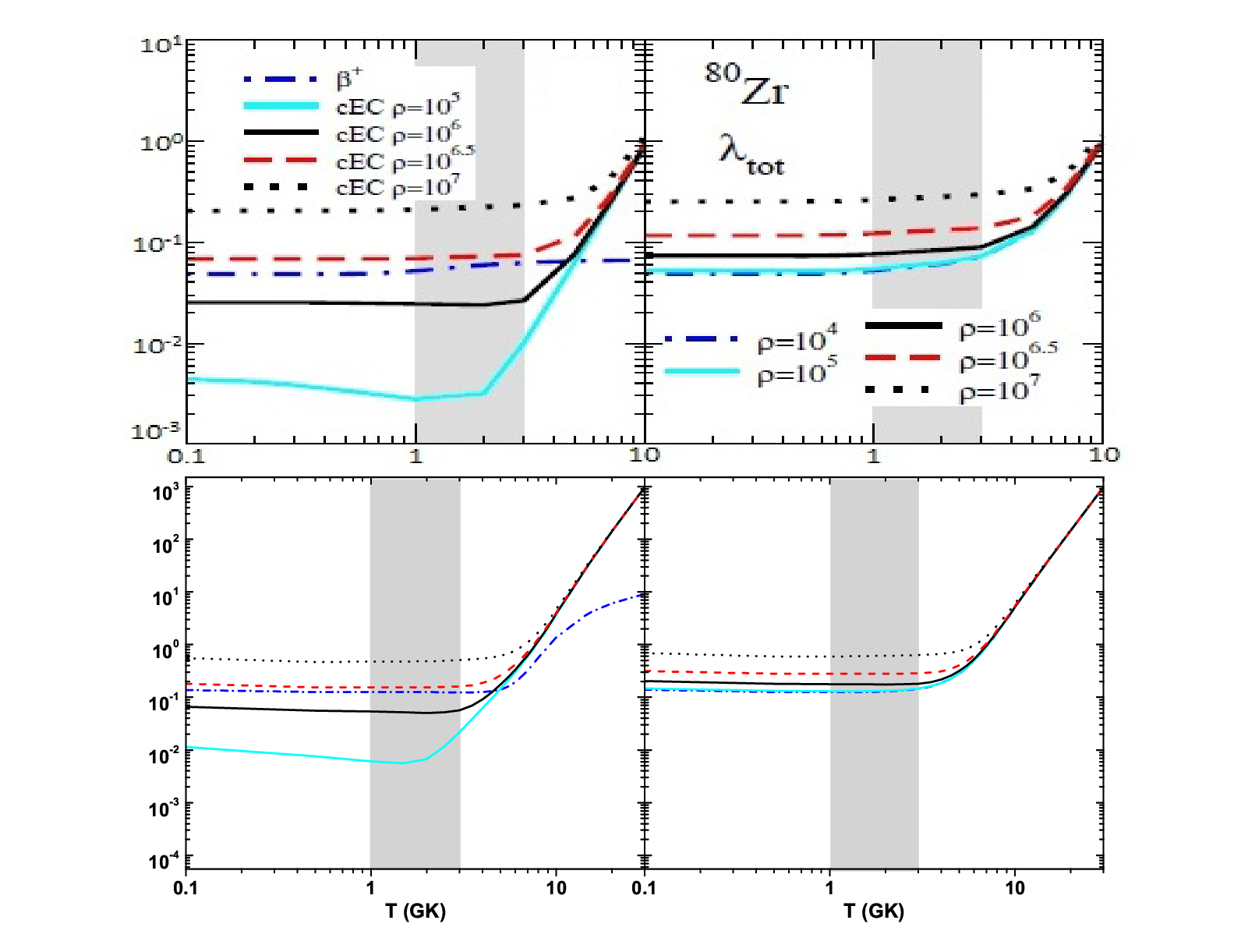}
\caption {(Color online) The comparison of the calculated
weak-interaction mediated rates for $^{80}$Zr as a function of
stellar temperature and density. The deformed Skyrme Hartree-Fock +
BCS  + QRPA calculation, reproduced from \cite{Sar11}, is shown in
the upper panels. The reported \mbox{pn-QRPA} calculation is shown
in the bottom panels. The electron capture and the positron decay
rates are shown in the left panels. In the right panels, the
combined total rates are shown. Units of all rates, the densities
and the temperatures are $s^{-1}$, $gcm^{-3}$ and $10^{9}$ K,
respectively.} \label{80Zr}
\end{figure}
\clearpage
\begin{figure}
\includegraphics[width=4.3in,height=4.3in]{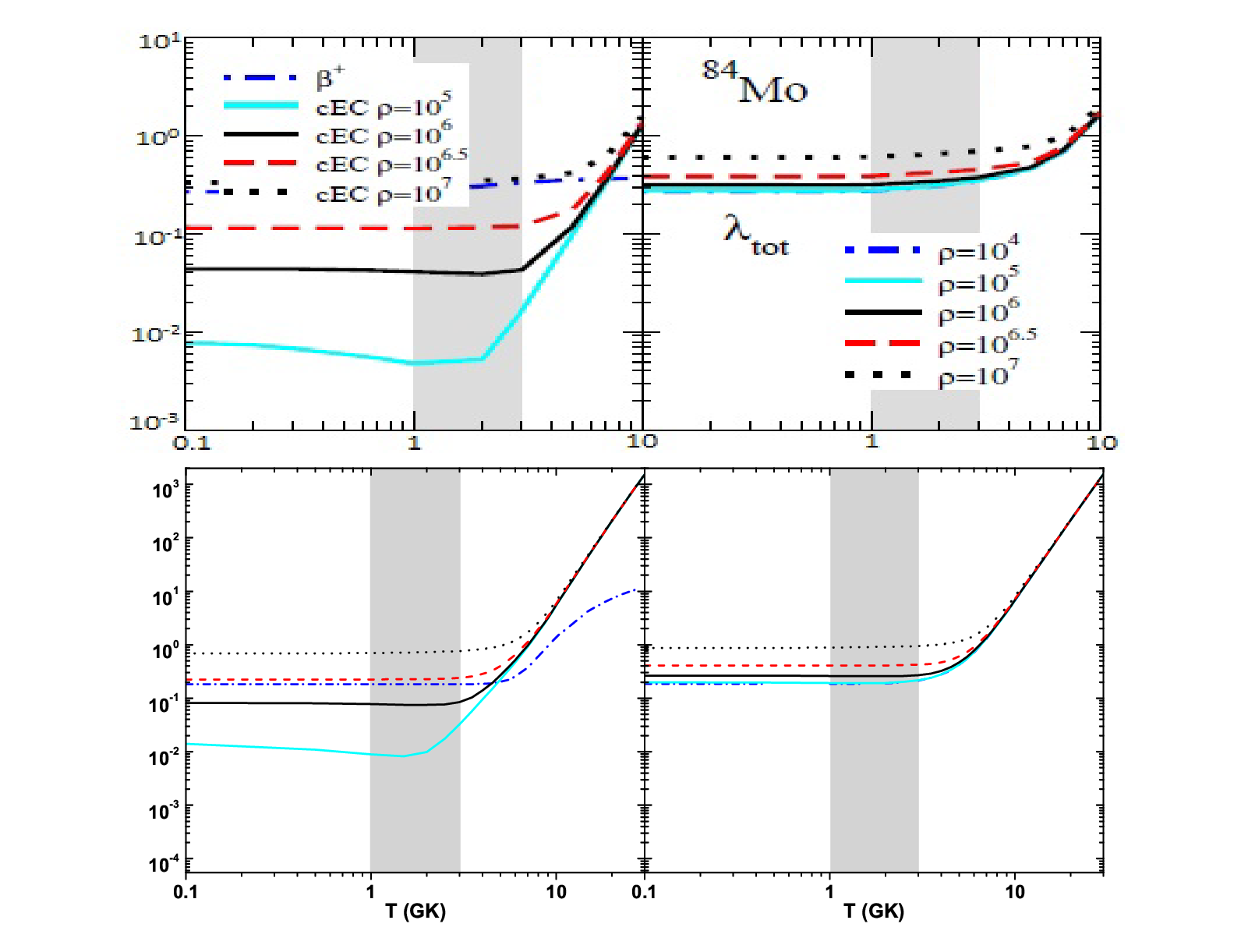}
\caption {(Color online) Same as Fig.~\ref{80Zr} but for $^{84}$Mo.}
\label{84Mo}
\end{figure}
\clearpage
\begin{figure}
\includegraphics[width=4.3in,height=4.3in]{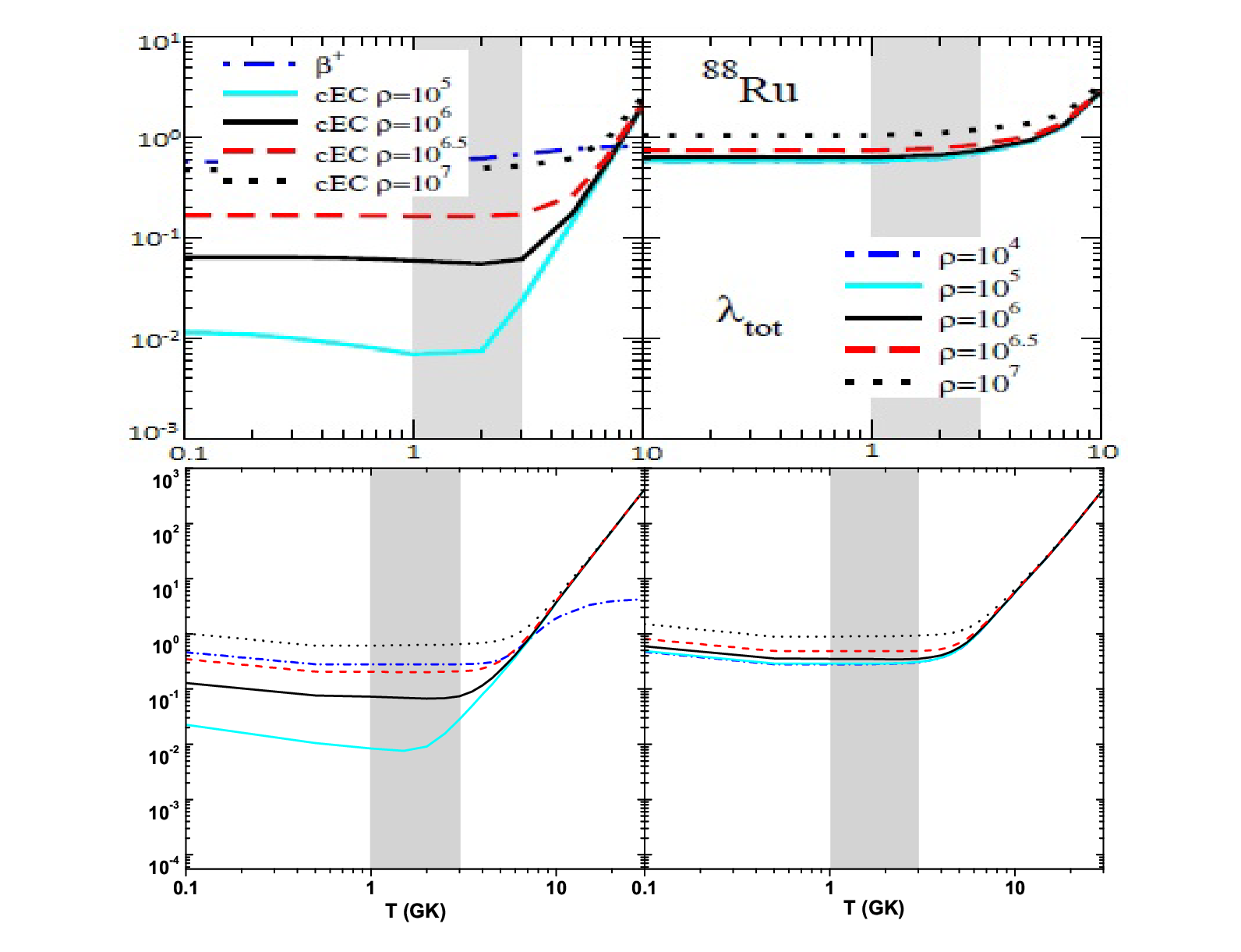}
\caption {(Color online) Same as Fig.~\ref{80Zr} but for $^{88}$Ru.}
\label{88Ru}
\end{figure}
\clearpage
\begin{figure}
\includegraphics[width=4.3in,height=4.3in]{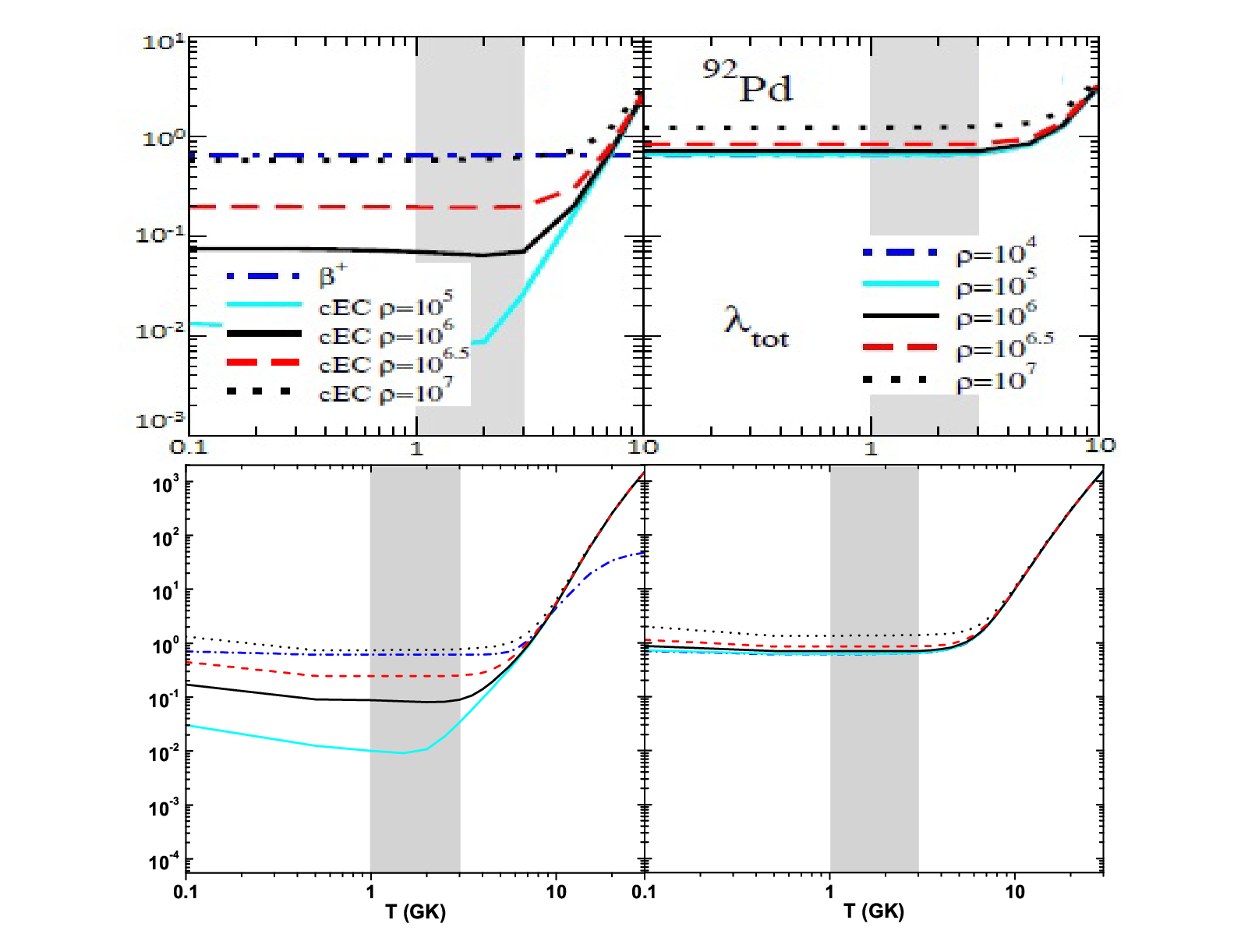}
\caption {(Color online) Same as Fig.~\ref{80Zr} but for $^{92}$Pd.}
\label{92Pd}
\end{figure}
\clearpage
\begin{figure}
\includegraphics[width=4.3in,height=4.3in]{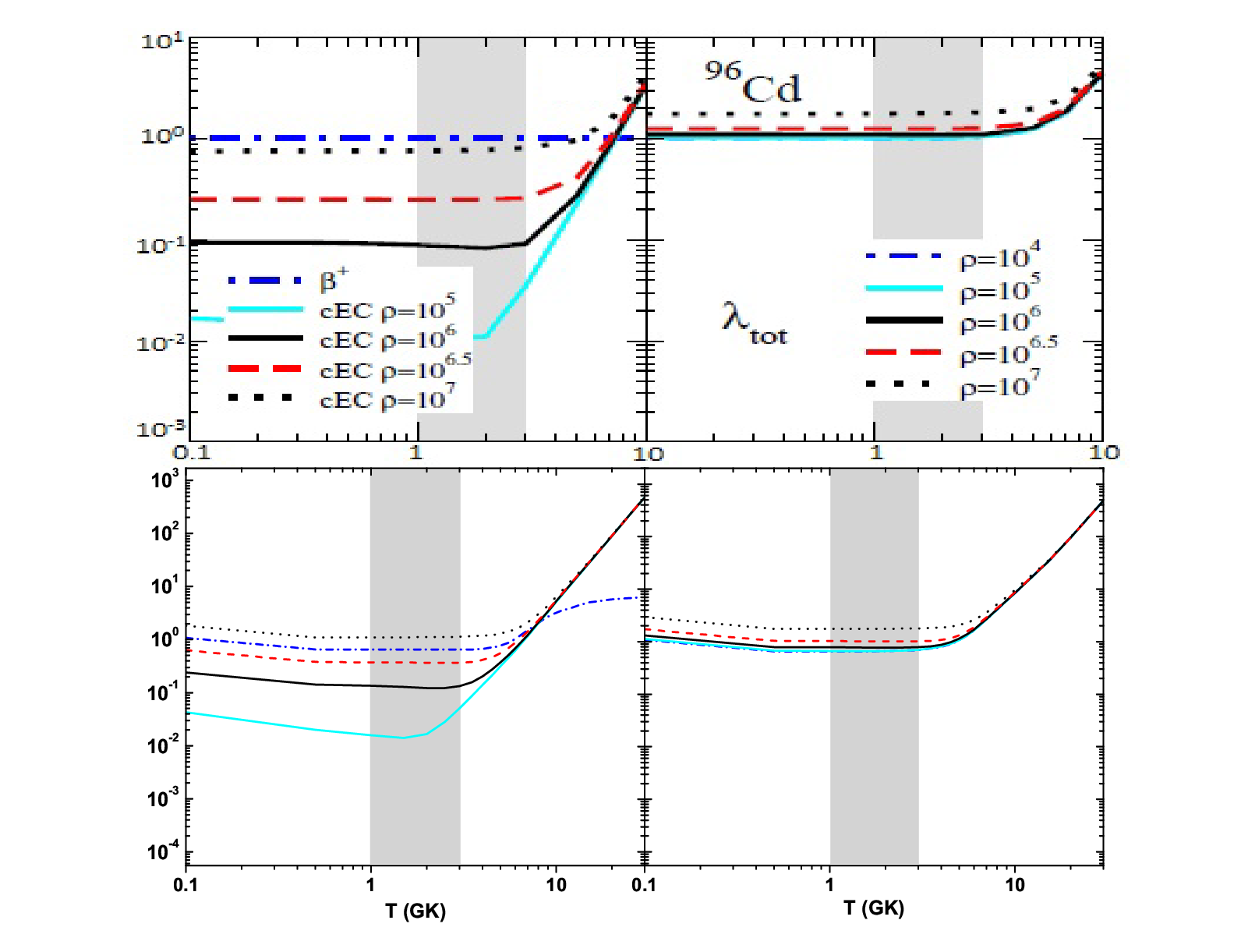}
\caption {(Color online) Same as Fig.~\ref{80Zr} but for $^{96}$Cd.}
\label{96Cd}
\end{figure}

\clearpage
\begin{table}
\begin{center}
\centering \caption{Deformation parameters calculated using the
IBM-1 Model (this work) and the RMF model~\cite{Lal99}.}\label{ta1}
\footnotesize
\begin{tabular}{@{}llllll}

~~~Nucleus~~~~&~~~~$^{80}$Zr~~~~&~~~~$^{84}$Mo~~~~&~~~~$^{88}$Ru~~~~&~~~~$^{92}$Pd~~~~&~~~~$^{96}$Cd~~~~\\
\hline
~~~~$\beta_{IBM}$&~~~~0.87~~~~&~~~~0.50~~~~&~~~~0~~~~&~~~~0~~~~&~~~~0~~~~\\
~~~~$\beta_{RMF}$~\cite{Lal99} &~~~~0.437 &~~~-0.247 &~~~~0.107 &~~~~0.112 &~~~~0.003\\
\hline
\end{tabular}
\end{center}
\end{table}

\begin{table}
\begin{center}
\centering \caption{IBM-1 Hamiltonian's parameters. $N$ is number of bosons, $\overline{\chi}$ is dimensionless and
others parameters have units of keV.}\label{ta2} \footnotesize
\begin{tabular}{@{}llllll}

~~~~~~~~~~~~~~~&~~~~$N$~~~~~&~~~~$\epsilon$~~~~&~~~~$a_{2}$~~~~&~~~~$\overline{\chi}$~~~~&~~~~$\sigma^{a}$~~~~~ \\
\hline
~~$^{80}$Zr~~ & ~~~~10~~~ & ~~~~314.8~~~~ & ~~~~-80~~~~~~ & ~~~~-0.25~~~~ & ~~~~6~~~~ \\
~~$^{84}$Mo~~ & ~~~~8~~~~ & ~~~~803.1~~~~ & ~~~~-48.5~~~~ & ~~~~-0.4~~~~~ & ~~~~8~~~~ \\
~~$^{88}$Ru~~ & ~~~~6~~~~ & ~~~~807.7~~~~ & ~~~~-41.3~~~~ & ~~~~-0.35~~~~ & ~~~~3~~~~ \\
~~$^{92}$Pd~~ & ~~~~4~~~~ & ~~~~871.2~~~~ & ~~~~~14.8~~~~ & ~~~~-0.35~~~~ & ~~~~26~~~ \\
~~$^{96}$Cd~~ & ~~~~2~~~~ & ~~~~1029.2~~~ & ~~~~~50~~~~~~ & ~~~~-0.35~~~~ & ~~~~0~~~~ \\
\hline
\end{tabular}\\
\end{center}
\end{table}

%

\begin{table}[htbp]
\begin{center}
\centering \caption{Calculated electron capture rates for allowed GT
transitions (in units of s$^{-1}$) and ratio of electron capture
rates to $\beta^{+}$-decay on
$^{80}$Zr,$^{84}$Mo,$^{88}$Ru,$^{92}$Pd and $^{96}$Cd at stellar
density of 10$^{6}$g/cm$^{3}$ using the \mbox{pn-QRPA}(N) model.
T$_{9}$ represents the temperature in 10$^{9}$K.} \label{ta4} \tiny
\hspace{0.5in}
\begin{tabular}{c|c|c|c|c|c|c|c|c|c|c}

 & \multicolumn{2}{|c|}{$^{80}$Zr} & \multicolumn{2}{|c|}{$^{84}$Mo} & \multicolumn{2}{|c|}{$^{88}$Ru} & \multicolumn{2}{|c|}{$^{92}$Pd}& \multicolumn{2}{|c}{$^{96}$Cd} \\
\cline{2-11}
T$_{9}$ & $\lambda_{cec}$ & R(cec/$\beta^{+}$) & $\lambda_{cec}$ & R(cec/$\beta^{+}$) & $\lambda_{cec}$ & R(cec/$\beta^{+}$) & $\lambda_{cec}$ & R(cec/$\beta^{+}$) & $\lambda_{cec}$ & R(cec/$\beta^{+}$) \\

\hline
0.5    & 5.5$\times10^{-2}$ & 4.4$\times10^{-1}$ & 8.1$\times10^{-2}$ & 4.4$\times10^{-1}$  & 7.6$\times10^{-2}$ & 2.7$\times10^{-1}$  & 9.1$\times10^{-2}$ & 1.5$\times10^{-1}$  & 1.4$\times10^{-1}$  & 2.2$\times10^{-1}$  \\
1.5    & 5.1$\times10^{-2}$ & 4.1$\times10^{-1}$ & 7.6$\times10^{-2}$ & 4.1$\times10^{-1}$  & 6.9$\times10^{-2}$ & 2.5$\times10^{-1}$  & 8.3$\times10^{-2}$ & 1.3$\times10^{-1}$  & 1.3$\times10^{-1}$  & 2.0$\times10^{-1}$  \\
2.5    & 5.1$\times10^{-2}$ & 4.1$\times10^{-1}$ & 7.6$\times10^{-2}$ & 4.2$\times10^{-1}$  & 6.8$\times10^{-2}$ & 2.4$\times10^{-1}$  & 8.1$\times10^{-2}$ & 1.3$\times10^{-1}$  & 1.2$\times10^{-1}$  & 1.9$\times10^{-1}$  \\
3.5    & 6.9$\times10^{-2}$ & 5.5$\times10^{-1}$ & 1.0$\times10^{-1}$ & 5.6$\times10^{-1}$  & 8.9$\times10^{-2}$ & 3.1$\times10^{-1}$  & 1.1$\times10^{-1}$ & 1.7$\times10^{-1}$  & 1.6$\times10^{-1}$  & 2.4$\times10^{-1}$  \\
4.5    & 1.3$\times10^{-1}$ & 9.7$\times10^{-1}$ & 1.9$\times10^{-1}$ & 1.0$\times10^{0}$   & 1.6$\times10^{-1}$ & 5.2$\times10^{-1}$  & 1.9$\times10^{-1}$ & 3.0$\times10^{-1}$  & 2.7$\times10^{-1}$  & 3.9$\times10^{-1}$  \\
5.5    & 2.4$\times10^{-1}$ & 1.5$\times10^{0}$  & 3.7$\times10^{-1}$ & 1.7$\times10^{0}$   & 3.0$\times10^{-1}$ & 7.6$\times10^{-1}$  & 3.6$\times10^{-1}$ & 5.1$\times10^{-1}$  & 5.1$\times10^{-1}$  & 5.8$\times10^{-1}$  \\
\hline

\end{tabular}
\end{center}
\end{table}

\end{document}